\documentclass[prb,twocolumn,aps,superscriptaddress,showpacs]{revtex4-2}

\usepackage{amsmath,amssymb}
\usepackage{graphicx}
\usepackage{xcolor}
\usepackage{graphicx}
\DeclareGraphicsExtensions{.pdf,.eps,.png,.jpg} \graphicspath{{./figs/}}
\usepackage{dcolumn}
\usepackage{bm}
\usepackage{tabularx}
\usepackage{booktabs}
\usepackage{upgreek}
\usepackage{multirow}
\usepackage{epstopdf}
\usepackage{gensymb}
\usepackage{comment}

\usepackage{amsmath}%
\usepackage{amsfonts}%
\usepackage{amssymb}%
\usepackage[latin1]{inputenc}
\usepackage{graphicx}
\usepackage{epstopdf}
\usepackage{color}
\usepackage{ulem}
\usepackage[english]{babel}
\usepackage{natbib}
\usepackage[colorlinks=true, citecolor=blue, linkcolor=blue, urlcolor=blue]{hyperref}
\usepackage{placeins}
\usepackage{booktabs}
\usepackage{enumitem}
\usepackage{subfigure}
\usepackage{multirow}
\usepackage{inputenc}
\usepackage{braket}
\usepackage{svg}
\usepackage{dcolumn}
\usepackage{bm}
\usepackage{siunitx}
\usepackage{mathtools}
\usepackage{soul}

\begin{document}

\title{Random singlet physics in the $S=\frac{1}{2}$ pyrochlore antiferromagnet NaCdCu$_2$F$_7$}

\author{A. Kancko}
\affiliation{Charles University, Faculty of Mathematics and Physics, Department of Condensed Matter Physics, Ke Karlovu 5, 121 16, Prague, Czech Republic}
\author{H. Sakai}
\affiliation{Advanced Science Research Center, Japan Atomic Energy Agency, Tokai, Ibaraki 319-1195, Japan}
\author{C. A. Corr\^ea}
\affiliation{Institute of Physics, Czech Academy of Sciences, Na Slovance 2, 182 00, Prague, Czech Republic}
\author{P. Proschek}
\affiliation{Charles University, Faculty of Mathematics and Physics, Department of Condensed Matter Physics, Ke Karlovu 5, 121 16, Prague, Czech Republic}
\author{J. Prokle\v{s}ka}
\affiliation{Charles University, Faculty of Mathematics and Physics, Department of Condensed Matter Physics, Ke Karlovu 5, 121 16, Prague, Czech Republic}
\author{T. Haidamak}
\affiliation{Charles University, Faculty of Mathematics and Physics, Department of Condensed Matter Physics, Ke Karlovu 5, 121 16, Prague, Czech Republic}
\author{M. Uhlarz}
\affiliation{Hochfeld-Magnetlabor Dresden (HLD-EMFL), Helmholtz-Zentrum Dresden-Rossendorf (HZDR), 01328
Dresden, Germany}
\author{A. Berlie}
\affiliation{ISIS Neutron and Muon Source, Rutherford Appleton Laboratory, Science and Technology Facilities Council, Chilton, Oxfordshire OX11 0QX, U.K.}
\author{Y. Tokunaga}
\affiliation{Advanced Science Research Center, Japan Atomic Energy Agency, Tokai, Ibaraki 319-1195, Japan}
\author{R. H. Colman}
\affiliation{Charles University, Faculty of Mathematics and Physics, Department of Condensed Matter Physics, Ke Karlovu 5, 121 16, Prague, Czech Republic}

\date{\today}

\begin{abstract}

We report a random singlet ground state in the $S=\frac{1}{2}$ Heisenberg pyrochlore antiferromagnet NaCdCu$_2$F$_7$. Cationic Na$^+$/Cd$^{2+}$ disorder on the pyrochlore $A$ site generates a broad distribution of Cu$^{2+}$--F$^-$--Cu$^{2+}$ exchange couplings, introducing intrinsic magnetic bond disorder. Despite strong antiferromagnetic interactions ($\theta_{\mathrm{CW}}=-72$~K), no magnetic order or global spin freezing is observed in DC and AC susceptibility, specific heat or $^{23}$Na nuclear magnetic resonance to 120 mK, with muon spin relaxation experiments confirming persistent spin dynamics to 58 mK. $T$-linear specific heat, a Curie-like susceptibility tail, and power-law scaling with data collapse in $\chi(T)$, $M(H)$, $C_{\mathrm{mag}}/T$, $^{23}$Na $(1/T_1T)$ and the muon spin polarization $P(t)$ reveal a disorder-driven network of random singlets and orphan spins. Scaling across multiple bulk and local probes is consistent with a broad distribution of exchange energies, $P[\mathcal{J}] \sim \mathcal{J}^{-\alpha}$. This behavior contrasts with previously-studied Na$A''B_2$F$_7$ pyrochlore fluorides, where magnetic bond disorder precipitates spin-glass freezing, underscoring the crucial role of strong $S=\frac{1}{2}$ quantum fluctuations in NaCdCu$_2$F$_7$.

\end{abstract}


\maketitle

\section{Introduction}

 Quantum spin liquids (QSL) -- highly entangled dynamic ground states without long-range magnetic order down to $T = 0$ -- remain a central focus of frustrated magnetism due to their exotic fractionalized excitations and non-trivial topology \cite{Balents2010,Savary2017,Zhou2017,Lancaster2023}, with possible applications in quantum technologies \cite{A.Yu.Kitaev2002,Nayak2008,Yang2018,Tokiwa2021}. The $S=\frac{1}{2}$ Heisenberg  antiferromagnet on the pyrochlore lattice is a canonical model \cite{Canals2000}: corner-sharing tetrahedra produce strong geometrical frustration, suppressing semiclassical order and promoting competing ground states, with theoretical predictions spanning chiral or dimerized phases and QSLs with short-range correlations \cite{Harris1991,Schafer2020,Hagymasi2021,Kim2008,Tsunetsugu2001,Astrakhantsev2021,Schneider2022,Chandra2018}.
 
Experimental realizations of ideal $S=\frac{1}{2}$ Heisenberg pyrochlore antiferromagnets are exceedingly rare. 
Multiple rare-earth $J_{\mathrm{eff}} = \frac{1}{2}$ pyrochlore oxides show QSL-like behaviour but suffer from strong anisotropy due to spin-orbit coupling and crystal field effects \cite{Gardner1999,Nakatsuji2006,Sibille2015},  whilst $J_{\mathrm{eff}} = \frac{1}{2}$ Co$^{2+}$-based pyrochlore fluorides freeze into glassy states due to magnetic bond disorder \cite{Kancko2023,Krizan2014,Krizan2015Co}. Lu$_2$Mo$_2$O$_5$N$_2$ \cite{Clark2014} and Hg$_2$Cu$_2$F$_6$S \cite{Kawabata2007} represent the only known realizations of true $S=\frac{1}{2}$ pyrochlores. Lu$_2$Mo$_2$O$_5$N$_2$ exhibits a dynamic, QSL-like ground state without spin freezing despite significant disorder, in stark contrast to its spin-glass $S=1$ counterpart Lu$_2$Mo$_2$O$_7$, underscoring the role of strong $S = \frac{1}{2}$ quantum fluctuations. Hg$_2$Cu$_2$F$_6$S also displays QSL-like behavior, but the lack of comprehensive low-temperature characterization leaves its ground-state nature unresolved.


In real materials, quenched disorder via site mixing, vacancies, or lattice distortions is unavoidable and can instead stabilize disorder-driven phases \cite{Ramirez2025}. One prominent outcome is the random singlet (RS) state \cite{Uematsu2019,Uematsu2017,Uematsu2018,Uematsu2021,Kawamura2014,Kawamura2019,Watanabe2014,Sanyal2021,Liu2018,Savary2017a,Shimokawa2025}, where a broad distribution of antiferromagnetic (AF) exchange energies produces localized singlets over multiple length scales, while weakly coupled moments form paramagnetic ``orphan spins''. The RS state can be distinguished from valence-bond solids (VBS) \cite{Isoda1998} and resonating valence-bond (RVB) manifolds \cite{Anderson1973} by its disorder-induced power-law distributions of exchange couplings $P[\mathcal{J}] \sim \mathcal{J}^{-\alpha}$ and density of states $\rho(E) \sim E^{-\alpha}$, leading to a characteristic $T$-linear specific heat, gapless susceptibility with a Curie-like tail, broad features in the dynamic spin structure factor, and power-law scaling of $\chi(T)$, $M(H)$, and $C_{\rm mag}/T$ with data collapse as a function of temperature and magnetic field $T/H$ \cite{Kimchi2018, Kimchi2018b, Dasgupta1980,Fisher1994,Bhatt1982}. RS behavior has been documented in 1D chains \cite{Volkov2020,Shiroka2019,Khatua2022}, 2D layered lattices \cite{Kundu2020,Kitagawa2018,Lee2023,Sheckelton2012,Kimchi2018b,Song2021,Yoon2021,Mustonen2018,Mustonen2024,Lee2024,Pal2020,Murayama2022,Do2014}, and a few 3D frustrated magnets \cite{Hossain2024,Sana2024}, but has not been established in a $S=\frac{1}{2}$ Heisenberg pyrochlore antiferromagnet.


In this work, we provide conclusive evidence for a RS ground state in NaCdCu$_2$F$_7$, making it the first $S = \frac{1}{2}$ Heisenberg pyrochlore where the RS state is confirmed across multiple probes. The inherent structural Na$^+$/Cd$^{2+}$ cationic disorder induces randomness in the Cu$^{2+}$-F$^-$-Cu$^{2+}$ magnetic superexchange pathways, forming the necessary distribution of exchange energies $\mathcal{J}$. Despite the strong AF interactions ($\theta_{\mathrm{CW}}= -72$ K), no magnetic transition or global spin freezing is seen in DC magnetization, AC susceptibility, specific heat or $^{23}$Na nuclear magnetic resonance (NMR), and we confirm continued dynamics down to 58 mK via zero field (ZF) and longitudinal field (LF) muon spin relaxation ($\mu$SR) measurements. Convincingly, evidence of power-law scaling and data collapse can be found in each of these experimental probes. The strong  quantum fluctuations of the $S =\frac{1}{2}$ Cu$^{2+}$ ions suppress the spin-glass freezing observed in related pyrochlore fluoride materials \cite{Kancko2023,Kancko2025,Krizan2014,Krizan2015Co,Krizan2015Ni,Sanders2017}, instead giving rise to a RS state. This establishes NaCdCu$_2$F$_7$ as a benchmark platform for testing theoretical predictions of 3D random singlet physics and exploring low-energy excitations, field-tuned phenomena, and exotic quantum spin-liquid behavior in three dimensions.

\begin{figure}[h]
\begin{center}
\includegraphics[scale=1]{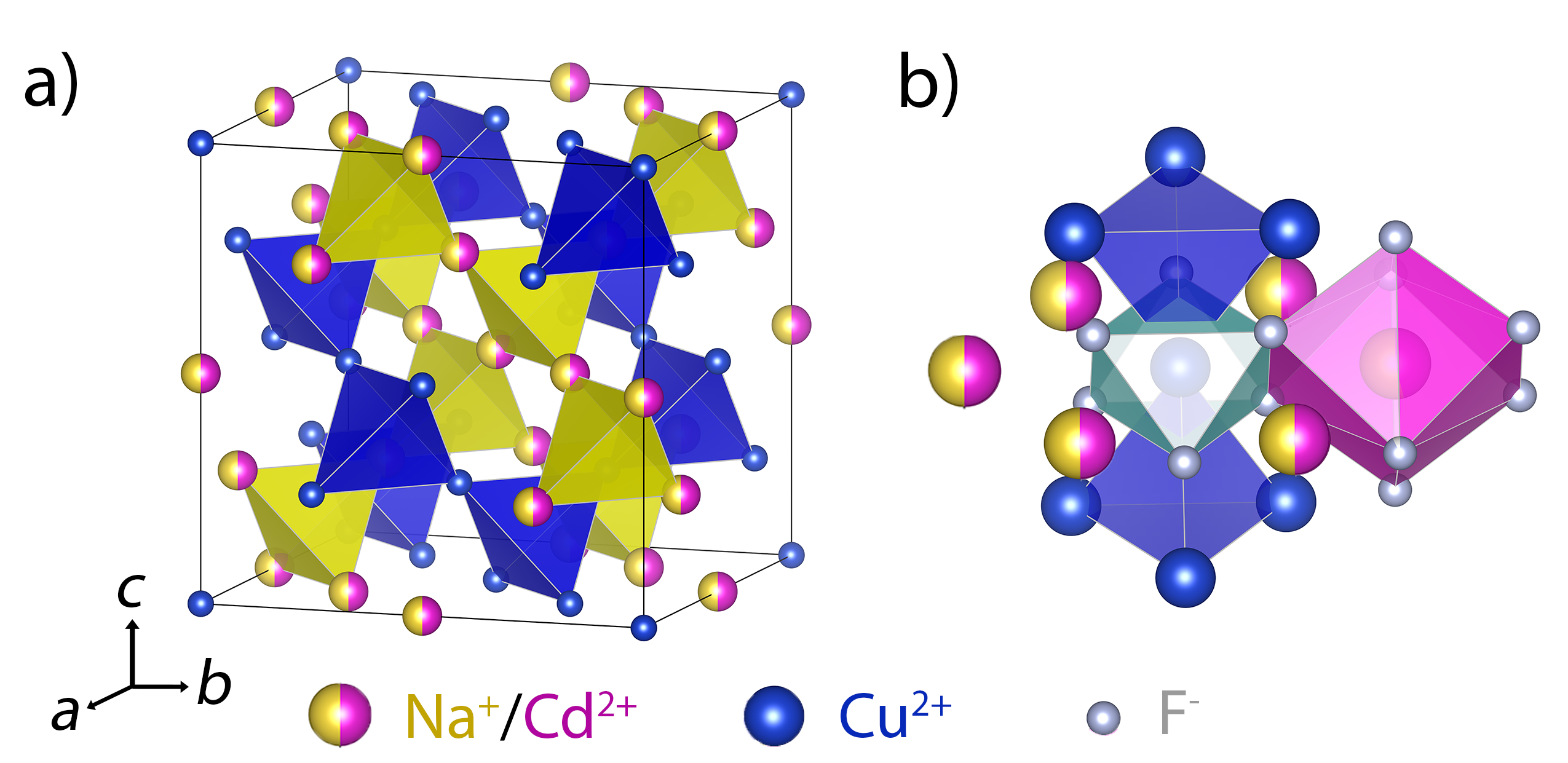}
\caption{(a) Cubic unit cell of NaCdCu$_2$F$_7$ with highlighted magnetic (blue) and non-magnetic (yellow) pyrochlore sublattices. Fluorine ions are ommitted for clarity. (b) Local environment around the magnetic Cu$^{2+}$ site - distorted CuF$_6$ octahedron (light blue), connected to a distorted (Na/Cd)F$_8$ cube (pink dodecahedron).} 
\label{fig1} 
\end{center}
\end{figure}

\section{Experimental methods}

 NaCdCu$_2$F$_7$ was synthesized using the optical floating-zone method by melting a stoichiometric fluoride mix in a hollow graphite crucible, yielding a largely polycrystalline rod. Phase purity was verified by powder X-ray diffraction (PXRD). Upon crushing the as-grown rod, small crystallites were isolated and used for structure solution using single crystal X-ray diffraction (SCXRD).

SCXRD was measured at 300 K using a Rigaku Xcalibur Gemini ultra-diffractometer, with graphite-monochromated Mo K$\alpha$ ($\lambda$ = 0.71073 \AA) radiation, and an Atlas S2 CCD detector. Diffraction data were integrated using CrysAlis Pro \cite{Crysalis} with an analytical numeric absorption correction combined with a multifaceted crystal model and an empirical absorption correction using spherical harmonics. The structure was solved by charge flipping using Superflip \cite{Palatinus2007} and refined by full-matrix least squares in Jana2020 \cite{Petricek2023}. Structural graphics were created using Vesta \cite{Momma2008}.

DC magnetization measurements on powdered NaCdCu$_2$F$_7$ were performed in a Quantum Design Magnetic Property Measurement System (model MPMS-XL 7 T), in the temperature range $T = 1.8\!-\!300$ K and the magnetic field range $\mu_0H = 0\!-\!7$ T, using the reciprocating sample option (RSO). 

High-field isothermal magnetization was measured in the High Magnetic Field Laboratory (Hochfeld-Magnetlabor Dresden, HLD) in the Helmholtz-Zentrum Dresden-Rossendorf (HZDR) in Dresden, Germany, using a pulsed 60 T magnet with a cryostat setup allowing the temperature range $T = 1.35\!-\!50$ K. 

AC susceptibility was measured in the frequency range $\nu_{\rm AC} = 215\!-\!10~ 000$ Hz between $T = 0.4\!-\!4$ K, using a custom-made coil for the $^3$He option in a Quantum Design Physical Property Measurement System (PPMS). 

Specific heat measurements were performed via the heat relaxation method in the PPMS between $T = 1.8\!-\!300$ K, with the addition of a $^3$He low-temperature insert for temperatures $T = 0.4\!-\!20$ K, in fields of $\mu_0H = 0\!-\!9$ T. The lattice specific heat was removed by subtracting the specific heat of the non-magnetic analogue NaCdZn$_2$F$_7$ from the total specific heat. \cite{Kancko2023,Kancko2025} An Oxford Instruments Triton dilution refrigerator (DR) was employed for zero-field and in-field heat capacity measurements between $T = 150\!-\!650$ mK.

ZF-$\mu$SR and LF-$\mu$SR measurements were performed at the ISIS Neutron and Muon Source at the Rutherford Appleton Laboratory in Harwell Oxford, United Kingdom. \cite{Colman2024} The MuSR instrument was used, utilising a dilution refrigerator cryostat setup for temperatures between $T = 0.05\!-\!4$ K, allowing applied longitudinal fields between $\mu_0H_{\rm LF} = 0\!-\!3000$ G. The obtained spectra were analyzed using WiMDA \cite{Pratt2000} and Mantid \cite{Arnold2014}.

NMR measurements were performed using a phase-coherent pulsed spectrometer with a superconducting NMR magnet. A $^4$He variable temperature insert (VTI) was utilized between $T = 1.8\!-\!150$ K, and a $^3$He-$^4$He dilution-refrigerator probe head was used between $T = 0.12\!-\!2.1$ K. A custom-wound  excitation coil from silver wire was used. Frequency-swept spectra at different constant fields were collected, with the radio-frequency (r.f.) circuit tuned and matched at each point. Nuclear spin echoes were generated using a standard 90$\degree$--180$\degree$ pulse sequence, with a first-pulse duration of 2 - 3 $\mu$s, where the r.f. power for nuclear spin excitation was optimized at each NMR spectral peak. The separation $\tau$ between the first and second pulse was typically 20 - 30 $\mu$s. Echoes were accumulated and Fourier transformed to obtain the spectra.

Details of the synthesis process, extended measurement details, and additional results of powder and single crystal X-ray difraction, magnetization, AC susceptibility, heat capacity, $^{23}$Na NMR, and $\mu$SR measurements can be found in the Supplementary Material (SM) \cite{Supplementary}. 

\section{Results}

\subsection{Structural details}

Structural studies confirmed the expected cubic pyrochlore structure (space group $Fd\overline{3}m$, $a$ = 10.3038(15) \AA) for NaCdCu$_2$F$_7$, depicted in Figure \ref{fig1}a, agreeing with the previously studied Na$A''B_2$F$_7$ fluorides \cite{Hansler1970, Kancko2023,Kancko2025,Krizan2014, Krizan2015Co,Krizan2015Ni,Sanders2017,Kubel2001,Oliveira2004,Grzechnik2007}. Structural randomness is inherent due to a mixed occupancy of Na$^+$/Cd$^{2+}$ cations fully disordered on the pyrochlore ($16d$) $A$ site. Inspection of the reciprocal space map cuts along the $(hk0), (h0l)$ and $(0kl)$ planes reveals no superstructure reflections or obvious diffuse scattering, indicating the absence of short- or long-range occupational ordering of Na$^+$ and Cd$^{2+}$ (see Figure S2 in SM \cite{Supplementary}).
The magnetic Cu$^{2+}$ ions on the pyrochlore ($16c$) $B$ site are octahedrally coordinated by F(2) fluorine on the $48f$ site, and are surrounded by six Na$^+$/Cd$^{2+}$ ions creating a hexagon in the (111) plane, see Figure \ref{fig1}b.  The difference in ionic radii of Na$^+$ (1.18 \AA) and Cd$^{2+}$ (1.10 \AA) \cite{Shannon1976}, in combination with their random occupation of the $A$ site, induce small trigonal distortions of the CuF$_6$ octahedra along the local [111] direction, which leads to a distribution of Cu$^{2+}$--F$^-$--Cu$^{2+}$ superexchange pathways - intrinsic magnetic bond disorder. The only free coordinate in the pyrochlore $Fd\overline{3}m$ space group, the F(2) fluorine $x$-coordinate $x_{\rm F(2)}$, dictates the size of the distortion of the CuF$_6$ octahedron as well as the (Na/Cd)F$_8$ dodecahedron. \cite{Kancko2023} The refined value, $x_{\rm F(2)} = 0.3316(4)$, results in a deformed CuF$_6$ octahedron with six equal Cu-F(2) bond lengths $d_{\rm Cu-F(2)} = 2.0060(17)$ \AA, while the F(2)-Cu-F(2) angle deviates from the ideal  90$\degree$ value by $\pm \delta = 7.22(10) \degree$. 

Refinement of the Cd occupancy, constrained with the Na occupancy by occ(Na)+occ(Cd) = 1, lead to a better solution than the ideal stoichiometric 0.5/0.5 case. A small Cd deficiency of occ(Cd) = 0.462(4) was obtained. To preserve charge balance, a refinement of the F(2) occupancy as a function of the Cd occupancy was attempted in the form of Na$_{1+x}$Cd$_{1-x}$Cu$_2$F$_{7-x}$, leading to occ(F2) = 0.924(7) with slightly better residue values, giving the final refined composition Na$_{1.076}$Cd$_{0.924}$Cu$_{2}$F$_{6.924}$. However, this sligtly off-stoichiometric structure does neither change the bond lengths (see Figure S4 in SM \cite{Supplementary}), nor the angles defining the CuF$_6$ and (Na/Cd)F$_8$ polyhedra, within the 3$\sigma$ limit.  

Atomic positions and thermal displacement parameters are summarized in Table \ref{tab1} and \ref{tab2}. Detailed SCXRD structure solution results and additional details of the refinement process can be found in SM \cite{Supplementary}.

\begin{table}[h]
\caption{Atomic positions of Na$_{1.076}$Cd$_{0.924}$Cu$_{2}$F$_{6.924}$.}
\renewcommand{\arraystretch}{1.5}
\setlength{\tabcolsep}{4pt}
\scalebox{1}{
\begin{tabular}{ccccccc}
\hline
\multirow{2}{*}{Atom} & \multirow{2}{*}{Site} & \multirow{2}{*}{$x$} & \multirow{2}{*}{$y$} & \multirow{2}{*}{$z$} & \multirow{2}{*}{Occupancy} & \multirow{2}{*}{\begin{tabular}[c]{@{}c@{}}Site\\ symmetry\end{tabular}} \\
                      &                       &                    &                    &                    &                            &                                                                          \\ \hline
Na                    & 16$d$                 & $\frac{1}{2}$      & $\frac{1}{2}$      & $\frac{1}{2}$      & 0.538(4)                   & $\overline{3}m$                                                         \\
Cd                    & 16$d$                 & $\frac{1}{2}$      & $\frac{1}{2}$      & $\frac{1}{2}$      & 0.462(4)                   & $\overline{3}m$                                                         \\
Cu                    & 16$c$                 & 0                  & 0                  & 0                  & 1                          & $\overline{3}m$                                                         \\
F(1)                    & 8$b$                  & $\frac{3}{8}$      & $\frac{3}{8}$      & $\frac{3}{8}$      & 0.924(7)                   & $\overline{4}3m$                                                         \\
F(2)                    & 48$f$                 & 0.3316(4)          & $\frac{1}{8}$      & $\frac{1}{8}$      & 1                          & $2mm$                                                                   \\ \hline
\end{tabular}}
\label{tab1}
\end{table}

\begin{table}[h]
\caption{Anisotropic thermal displacement parameters of Na$_{1.076}$Cd$_{0.924}$Cu$_{2}$F$_{6.924}$ at $T=300$ K.}
\renewcommand{\arraystretch}{1.5}
\setlength{\tabcolsep}{4pt}
\scalebox{0.9}{
\begin{tabular}{cccccccclcccc}
\hline
\multirow{2}{*}{Atom} & \multicolumn{12}{c}{$U_{\rm ij}$ ($\times 10^{-4}$ \AA$^2$)}                                                     \\ \cline{2-13} 
                      & $U_{11}$ & $U_{22}$ & $U_{33}$ &          & $U_{12}$        & $U_{13}$  & \multicolumn{2}{c}{} & $U_{23}$ &  & $U_{\rm ani}$ &  \\ \hline
Na                    & 258(7)   & $U_{11}$ & $U_{11}$ & \multicolumn{2}{c}{-51(3)} & $U_{12}$  & \multicolumn{4}{c}{$U_{12}$}       & 258(4)        &  \\
Cd                    & 258(7)   & $U_{11}$ & $U_{11}$ & \multicolumn{2}{c}{-51(3)} & $U_{12}$  & \multicolumn{4}{c}{$U_{12}$}       & 258(4)        &  \\
Cu                    & 165(6)   & $U_{11}$ & $U_{11}$ & \multicolumn{2}{c}{17(3)}  & -$U_{12}$ & \multicolumn{4}{c}{-$U_{12}$}      & 165(3)        &  \\
F(1)                    & 151(18)  & $U_{11}$ & $U_{11}$ & \multicolumn{2}{c}{0}      & 0         & \multicolumn{4}{c}{0}              & 151(10)       &  \\
F(2)                    & 470(20)  & 585(18)  & $U_{22}$ & \multicolumn{2}{c}{0}      & 0         & \multicolumn{4}{c}{60(20)}         & 546(12)       &  \\ \hline
\end{tabular}}
\label{tab2}
\end{table}

\subsection{Magnetic properties}

\begin{figure}[h]
\begin{center}
\includegraphics[scale=1.1]{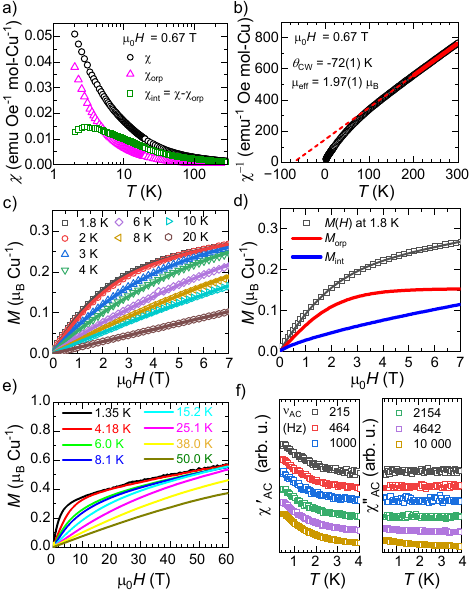}
\caption{(a) Magnetic susceptibility $\chi(T)$ at $\mu_0 H$ = 0.67 T, decomposed into an orphan spin contribution $\chi_{\rm orp}$ and intrinsic susceptibility $\chi_{\rm int} = \chi - \chi_{\rm orp}$. (b) The inverse total susceptibility $\chi^{-1}(T)$, fitted to a Curie-Weiss law between 200 and 300 K. (c) Isothermal magnetization $M(H)$ between 1.8 and 20 K, with fits to Equation \ref{eq1} shown as solid lines. (d) Decomposition of the 1.8 K $M(H)$ curve into the fitted orphan spin $M_{\rm orp}$ and intrinsic $M_{\rm int}$ terms. (e) High-field magnetization between 1.35 and 50 K. (f) Real part (left) and imaginary part (right) of AC magnetic susceptibility, vertically offset for clarity.} 
\label{fig2} 
\end{center}
\end{figure}

Temperature-dependent magnetic susceptibility, $\chi(T) = M(T)/H$, shows no magnetic transition down to 1.8 K, seen in Figure \ref{fig2}a. Instead a diverging Curie-like tail is noted below 10 K, a typical sign of paramagnetic ``orphan'' spins dominating the low-temperature bulk magnetic properties. The orphan spin response, $n_{\rm orp} \sim 15 \%$ was identified and subtracted by fitting the isothermal magnetization $M(H)$ at 1.8 K (see detailed analysis below), resulting in the intrinsic susceptibility, $\chi_{\rm int} = \chi - \chi_{\rm orp}$, also shown in Figure \ref{fig2}a. The  $\chi_{\rm int}$ displays a broad maximum centered around 3 K. A Curie-Weiss fit of the inverse total susceptibility $\chi^{-1}(T)$ between 200--300 K, shown in Figure \ref{fig2}b, gives $C$ = 0.486(1) emu K mol$^{-1}$ Oe$^{-1}$ and  $\theta_{\mathrm{CW}} = -72(1)$ K, resulting in a $\mu_{\mathrm{eff}}$ = 1.97(1) $\mu_B$ \cite{Mugiraneza2022}. The large negative $\theta_{\mathrm{CW}}$ confirms dominant AF interactions, and using $\theta_{\rm CW} = 2zS(S + 1)\mathcal{J}/3$ with $z = 6$ the nearest-neighbor coordination number of the pyrochlore lattice, we estimate the exchange coupling constant $\mathcal{J} = -24$ K. The $\chi_{\rm int}$ qualitatively follows recent theoretical predictions, although the observed broad maximum at $T\approx0.12\mathcal{J}$ is notably lower in temperature than that predicted by numerical linked cluster expansion (NLCE) results for the disorder free system, at $T=0.54\mathcal{J}$ \cite{Schafer2020}. The moment size of $\mu_{\mathrm{eff}}=1.97(1)$ $\mu_B$ is consistent with other Cu$^{2+}$ systems \cite{Kundu2020,Khatua2022, Mustonen2018}, and close to the theoretical spin-only moment $\mu^{\rm theo}_{\mathrm{eff}} = g\sqrt{J(J+1)} \mu_B$ = 1.73 $\mu_B$, for $J=S=\frac{1}{2}$ and $g=2$ with quenched orbital momentum.


Isothermal magnetization $M(H)$, seen in Figure \ref{fig2}c, shows a linear AF response at 20 K, and begins to show a Brillouin-like curvature for $T \leq 10$ K at fields below 5 T, superimposed on a linear response visible above 5 T. The $M(H)$ data are well fitted by a superposition of a modified $S=\frac{1}{2}$ Brillouin function corresponding to weakly interacting orphan spins $M_{\rm orp} = n_{\rm orp}B_{J=1/2}(x')$, and an intrinsic power-law term $M_{\rm int} = A(\mu_0H)^\alpha$ \cite{Lee2024,Sana2024,Do2014}:
\begin{equation}
 M(H) = \frac{n_\mathrm{orp} g \mu_B}{2} \mathrm{tanh} \left(\frac{g \mu_B \mu_0 H}{2k_\mathrm{B} (T-T^*)}\right) + A(\mu_0 H)^\alpha
 \label{eq1}
\end{equation}
where $n_\mathrm{orp}$ is the orphan spins fraction, $g$ is the orphan-spin $g$-factor fixed to $g=2$, $T^*$ is the interaction temperature between the orphan spins and $\alpha$ is the power-law exponent. From the $M(H)$ fit at 1.8 K, seen in Figure \ref{fig2}d, we extract $n_{\rm orp}$ = 0.154(2), $T^*$ = 0.30(2) K and $\alpha$ = 0.73(1), corresponding to $\sim$15\% of weakly ferromagnetically-coupled orphan spins. This $n_{\rm orp}$ value was used in the $\chi_{\rm int}=\chi-\chi_{\rm orp}$ subtraction in Figure \ref{fig2}a. From the 1.8 K $M(H)$ decomposition into $M_{\rm orp}$ and $M_{\rm int}$ shown in Figure \ref{fig2}d, we observe a full saturation of the orphan-spin component by $\mu_0H \sim  5$ T. The sublinear behaviour $\alpha < 1$ of the intrinsic AF response comes about as a result of the disorder-induced distribution of exchange energies. With this orphan spin fraction $n_{\rm orp}$ and the  interaction temperature $T^*$ fixed to that from the lowest temperature data set where it is best defined, we performed a global fit for $M(H)$ between 2 and 20 K. The fits are shown as solid lines in Figure \ref{fig2}c. The power-law exponent $\alpha$ varied between 0.68 and 0.99, indicating a crossover from Brillouin-like to purely linear ($\alpha \sim 1$) magnetization at high temperature (see Table S2 in SM \cite{Supplementary}).  

High-field isothermal magnetization $M(H)$ in pulsed field up to 60 T, seen in Figure \ref{fig2}e at temperatures between 1.35--50 K, exhibits no plateaus or field-induced transitions predicted for an ideal, disorder-free $S=\frac{1}{2}$ Heisenberg pyrochlore antiferromagnet \cite{Pal2019,Hagymasi2022}. Instead, a rapid low-field ($\mu_0H <$ 5 T) polarization of weakly coupled orphan spins is observed at the lowest temperatures, followed by a slow linear polarization of dynamically fluctuating strongly coupled singlets up to 60 T. This behaviour highlights the broad distribution of exchange couplings expected from the RS state, and is consistent with previously-studied RS materials \cite{Volkov2020,Sana2024,Do2014}. Even at 60 T we reach only $\sim\!0.6$ $\mu_{\rm B}$/Cu at 1.35 K, remaining far below full saturation expected for a spin-only Cu$^{2+}$ ion ($\mu_S^{\mathrm{theo}}$ = $2S\mu_{\rm B}$ = 1 $\mu_{\rm B}$), underscoring the robustness of strongly bound singlets and persistent fluctuations. 

No bifurcation of zero-field-cooled, $\chi_{\rm ZFC}$,  and field-cooled, $\chi_{\rm FC}$, magnetic susceptibility measurements was observed down to 1.8 K, indicating no history dependence, in accordance with a dynamic state (see Figure S5 in SM \cite{Supplementary}). AC susceptibility, measured with an alternating field amplitude $H_{\mathrm{AC}}$ = 0.05 Oe at frequencies between 215 and 10 000 Hz, shows an increase with decreasing temperature without reaching a maximum in the real part, $\chi'_{\mathrm{AC}}(\nu,T)$, while the imaginary part $\chi''_{\mathrm{AC}}(\nu,T)$ remains flat without signs of energy-dissipative processes, see Figure \ref{fig2}f. No frequency dependence was seen in either the real or imaginary parts, ruling out any global collective spin freezing into a static spin-glass ground state down to 0.4 K, despite the sizable magnetic bond disorder present in this material. This is in direct contrast to all previously studied bond-disordered Na$A''M_2$F$_7$ pyrochlore fluorides ($A''$ = Cd, Ca, Sr; $M$ = Co, Ni, Mn, Fe), which all reported a glassy ground-state, and a frequency-dependent peak in $\chi'_{\mathrm{AC}}(\nu,T)$ at $T_{\mathrm{f}} \approx 2 - 4$  K. \cite{Kancko2023,Krizan2014,Krizan2015Co,Krizan2015Ni, Sanders2017,Kancko2025}

\subsection{Specific heat}

\begin{figure}
    \centering
    \includegraphics[width=1\linewidth]{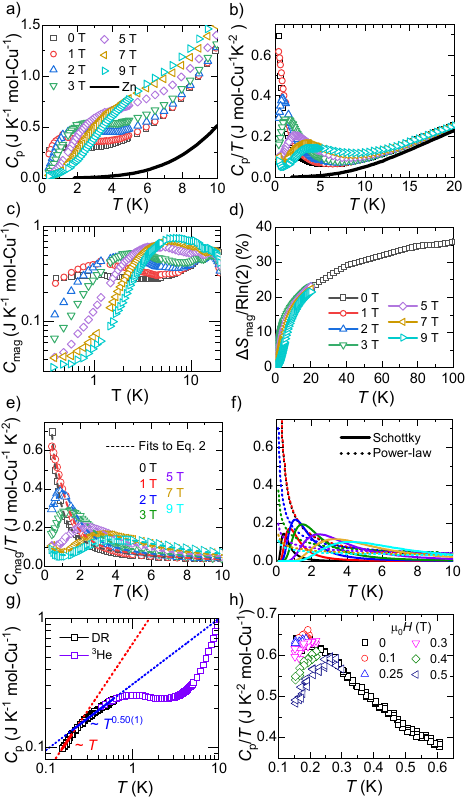}
    \caption{(a) Specific heat $C_p$ and (b) $C_p/T$ of NaCdCu$_2$F$_7$ in fields of 0--9 T, with the non-magnetic analogue NaCdZn$_2$F$_7$ shown as a black solid line. (c) Magnetic specific heat $C_{\rm mag}$ after subtraction of phonon contribution. (d) Magnetic entropy change $\Delta S_{\rm mag}$ in \% of Rln(2). (e) Fits of $C_{\rm mag}/T$ to Equation \ref{eq2}, with (e) a decomposition into the Schottky and power-law terms. (g) Zero-field dilution refrigerator (DR) and $^3$He specific heat $C_p$ with a power-law ($T = 200\!-\!650$ mK) and linear ($T<200$ mK) regime. (h) DR $C_p/T$ between 0--0.5 T.}
    \label{fig3}
\end{figure}

Specific heat, $C_p(T)$, in zero field shows no sharp first or second order transitions down to 150 mK, ruling out the development of long-range order or global structural changes, however a broad hump is observed at around 1 K, seen in Figure \ref{fig3}a and \ref{fig3}b. This feature shifts to higher temperature with applied field, reminiscent of a Schottky anomaly. After subtraction of the phonon contribution by subtracting the specific heat of the non-magnetic analogue NaCdZn$_2$F$_7$ \cite{Kancko2023,Kancko2025} (black solid line in Figure \ref{fig3}a and \ref{fig3}b) from the total $C_p$, the magnetic specific heat $C_{\mathrm{mag}} = C_p-C_{\rm lat} \approx C_{p\rm[NaCdCu_2F_7]}-C_{p\rm [NaCdZn_2F_7\rm]}$, and its evolution in applied fields is shown in Figure \ref{fig2}c. The magnetic entropy change $\Delta S_{\rm mag}$, calculated by integration of $C_{\mathrm{mag}}/T$ and shown in Figure \ref{fig3}d, reaches only $\sim\!35\%$ of the expected value of Rln2, by 100 K. The large missing entropy is a signature of non-magnetic singlet formation combined with residual dynamics at low temperature. 

$C_{\rm mag}(T,H)$ was fitted using a combination of a two-level Schottky anomaly $C_{\rm Sch}$ associated with weakly coupled $S=\frac{1}{2}$ orphan spins, and a power-law term $C_{\rm power-law} = A_{\rm power-law}T^{1-\alpha_{C_{mag}}}$ corresponding to the RS phase \cite{Kimchi2018,Murayama2022,Hossain2024}:

\begin{equation}
   \small \frac{C_{\rm mag}}{T} = A_{\rm Sch}\mathrm{R} \frac{\Delta^2}{T^3} \frac{\mathrm{exp}\left(\frac{\Delta}{T}\right)}{\left(1+\mathrm{exp}\left(\frac{\Delta}{T}\right) \right)^2} + A_{\rm power-law}T^{-\alpha_{C_{\rm mag}}}
    \label{eq2}
\end{equation}

The fits to Equation \ref{eq2} are summarized in Figure \ref{fig3}e, with the individual contributions shown in Figure \ref{fig3}f. We find that $\alpha_{C_{mag}}$ decreases from 0.87(2) in zero field to 0.28(2) at 9 T, and the extracted Schottky gap $\Delta$ in zero field is 1.34(5) K, suggesting a non-zero splitting of the Cu$^{2+}$ $S=\frac{1}{2}$ energy levels due to the local internal fields. From the Zeeman formula $\Delta(H) = gS\mu_B \mu_0 H$ we obtain the slope $gS = 1.32(6)$, giving the orphan-spin $g$-factor $g = 2.6(1)$ (see Figure S8 in SM \cite{Supplementary}). The orphan spins fraction $n_{\rm oph}$ can be estimated from the ratio of the Schottky entropy and total magnetic entropy $n_{\rm oph} = \Delta S_{\rm Sch}/\Delta S_{\rm mag}$, yielding an apparent increase from $\sim 10 \%$ in zero field to $\sim 54 \%$ at 9 T (see Table S4 in SM\cite{Supplementary}). This increase does not reflect the creation of new orphan spins. Rather, it arises from the field-induced polarization of weakly bound random singlet pairs, which contribute to the low-temperature Schottky-like feature only once the applied field overcomes their singlet binding energy. In zero field these pairs remain largely inactive, so the intrinsic orphan-spin density is field-independent and best represented by the zero-field value.

At temperatures between 200 and 650 mK, the specific heat shows a power-law dependence $C_p \sim T^{1-\alpha_{C_p}}$ with $\alpha_{C_p} = 0.50(1)$, followed by a $T$-linear specific heat $C_p = \gamma T$ below 200 mK, with $\gamma$ = 0.64(2) J K$^{-2}$ mol-Cu$^{-1}$, see Figure \ref{fig3}g. The crossover to the linear regime at the lowest temperatures is a characteristic thermodynamic signature of the RS state \cite{Uematsu2019,Uematsu2017,Uematsu2018,Uematsu2021,Kawamura2014,Kawamura2019}. In close analogy to the Sommerfeld coefficient in metals, the linear specific-heat coefficient $\gamma$ in QSL/RS Mott insulators is proportional to the density of low-energy magnetic excitations in the zero-energy limit, consistent with a gapless excitation spectrum. In Figure \ref{fig3}h, we observe a constant $C_p/T$ below 200 mK in zero field corresponding to the $C_p/T=\gamma$ regime, and an emerging sharp peak in applied 0.1--0.5 T fields, which shifts to higher temperature with increasing field. The emergence of a sharp low-temperature peak in $C_p/T$ under small applied magnetic fields may indicate the development of short-range correlations involving orphan spins, stabilized by the suppression of low-energy random singlet fluctuations. 

\subsection{Muon spin relaxation ($\mu$SR)}

\begin{figure}
    \centering
    \includegraphics[width=1\linewidth]{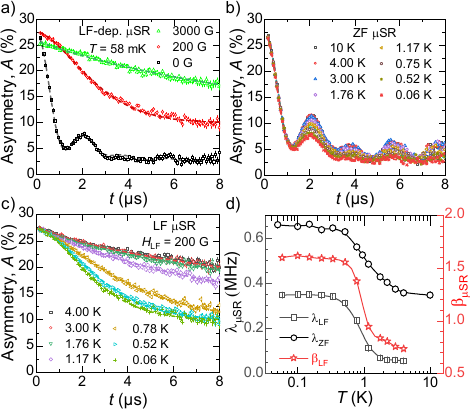}
    \caption{(a) LF-dependence of the $T=58$ mK spectra, showing the ZF F-$\mu$-F oscillations (black), stretched-exponential relaxation of asymmetry at 200 G (red) and a purely exponential relaxation at 3000 G (green). Temperature dependence of (b) the ZF spectra, fitted to Equation \ref{eq3} and (c) the LF 200 G spectra, fitted to Equation \ref{eq5}. (d) Temperature dependence of the fitted ZF and LF muon relaxation rates $\lambda_{\rm LF}$ and $\lambda_{\rm ZF}$ (left axis), and the stretching exponent $\beta_{\rm LF}$ (right axis).}
    \label{fig4}
\end{figure}

Spin dynamics were characterized down to 58 mK using local-probe $\mu$SR experiments. Zero-field measurements revealed oscillations coming from the F$^-$--$\mu^+$--F$^-$ bound complex, commonly seen in fluoride materials \cite{Brewer1986,Cai2018,Wilkinson2020}. The base-temperature ZF muon relaxation can be seen in Figure \ref{fig4}a, and the temperature dependence of ZF-$\mu$SR is shown Figure \ref{fig4}b.  The ZF spectra were fitted using a simple exponential function multiplied by a modified, phase-shifted ZF F-$\mu$-F function $G_{\rm F-\mu-F}(t)$:

\begin{equation}
    A_{\rm ZF}(t) = A_0 G_{\rm F-\mu-F}(t) \mathrm{exp}(-\lambda_{\rm ZF} t) + A_{\rm base} 
    \label{eq3}
\end{equation}

where $A_0$ is the relaxing amplitude, $\lambda_{\rm ZF}$ is the zero-field muon relaxation rate, $A_{\rm base}$ is a non-relaxing constant baseline and the ZF F-$\mu$-F function $G_{F-\mu-F}$ has the form \cite{Brewer1986}

\begin{align}
G_{\rm F-\mu-F}(t) = \frac{1}{6} \Big[ \;&
3 + \cos(\sqrt{3}\,\omega_d t + \phi) \nonumber\\
&+ \left( 1-\frac{1}{\sqrt{3}} \right)
\cos\!\left( \frac{3-\sqrt{3}}{2}\,\omega_d t + \phi \right) \nonumber\\
&+ \left( 1+\frac{1}{\sqrt{3}} \right)
\cos\!\left( \frac{3+\sqrt{3}}{2}\,\omega_d t + \phi \right)
\Big].
\label{eq4}
\end{align}

Here, $\phi$ is the phase shift and $\omega_d$ is the dipolar interaction frequency. On cooling from 10 K in zero field, the F$^-$--$\mu^+$--F$^-$ oscillations are damped by electronic fluctuations slowing into the time-window of the muon. This exponential damping increases significantly below $\sim\!3$ K, indicating a rapid slowing of the spin fluctuation rate, but plateaus below $\sim\!0.7$ K - a hallmark of continued quantum fluctuations \cite{Yang2024,Li2016}, see Figure \ref{fig4}d. This plateau is also seen in measurements in an applied 200 G longitudinal field, used to decouple the muons from the static dipolar fields of the F$^-$--$\mu^+$--F$^-$ state, seen in Figure \ref{fig4}d. 
However, the cleaner spectra, shown in Figure \ref{fig4}c, highlight a decay of asymmetry at low temperatures that is not purely exponential, but instead show a Gaussian shoulder at $t<3$ $\mu$s, and are better fitted by a stretched exponential function
\begin{equation}
    A_{\rm LF}(t) = A_0 \mathrm{exp}(-(\lambda_{\rm LF} t)^{\beta_{\rm LF}}) + A_{\rm base},
    \label{eq5}
\end{equation}

where $A_0$ is the relaxing amplitude, $\lambda_{\rm LF}$ is the LF muon relaxation rate, $\beta_{\rm LF}$ is the stretching exponent and $A_{\rm base}$ is a non-relaxing constant baseline. The stretched-exponential relaxation indicates an inhomogeneous mixture of both dynamic and partially frozen local fields. 
This can be explained by a slowing of dynamics of the weakly interacting orphan spins below 1 K within the RS matrix \cite{Uemura1994,DeVries2013,Sana2024}. The inhomogeneous distribution of local fields is tracked using the stretching exponent, $\beta_{\rm LF}$, in Figure \ref{fig4}d. At higher-longitudinal fields these weakly interacting frozen orphan spins are fully polarized and the spectra return to a purely exponential relaxation, shown in Figure \ref{fig4}a, confirming continued spin-dynamics even at 58 mK. 

\subsection{Nuclear magnetic resonance (NMR)}

\begin{figure}
    \centering
    \includegraphics[width=1\linewidth]{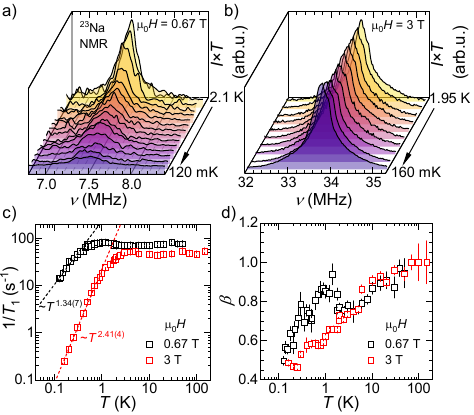}
    \caption{Temperature dependence of the $^{23}$NMR spectral intensity scaled by temperature $I \times T$ (a) at  $\mu_0H$ = 0.67 T and (b) at $\mu_0H$ = 3 T. Temperature dependence of (c) the $^{23}$Na spin-lattice relaxation rate $1/T_1$ and (d) the stretching exponent $\beta_{\rm NMR}$.}
    \label{fig5}
\end{figure}

To complement the $\mu$SR results, $^{23}$Na NMR measurements were performed between 120 mK and 150 K in fields of 0.67 T and 3 T. $^{23}$Na occupies the axially symmetric $A$ site (point group $\bar{3}m$), where it is statistically distributed with Cd$^{2+}$ with equal probability. As shown in Figure \ref{fig5}a and \ref{fig5}b, the spectra exhibit a single broad peak with broad tails on both sides. The spectral shape likely reflects a broadened $I\!=\!\frac{3}{2}$ powder pattern, in which the Na/Cd site distribution merges the central and quadrupolar satellite transitions. In the related compound NaCaCo$_2$F$_7$, the $^{23}$Na nuclear quadrupole frequency was reported to be of the order of $\nu_Q$($^{23}$Na) $\sim$ 1~MHz \cite{Sarkar2017}. At every temperature, the nuclear spin-lattice relaxation rate $1/T_1$ was measured at the position of the resonance peak by fitting the recovery curves with a stretched exponential function:
\begin{align}
    \frac{M(\infty)-M(t)}{M(\infty)} = 0.4 \exp \{-\left( t/T_1 \right)^{\beta_{\rm NMR}} \} \nonumber \\
    + 0.6 \exp \{ -\left( 6t/T_1 \right)^{\beta_{\rm NMR}} \},
    \label{eq6}
\end{align} 

assuming that only the central transition was saturated while the populations of satellite transitions remain unaffected. The temperature dependence of the extracted $1/T_1$ and $\beta_{\rm NMR}$ values is summarized in Figure \ref{fig5}c and \ref{fig5}d.

The stretching exponent $\beta_{\rm NMR}$ relates to the spatial distribution of $T_1$, with $\beta_{\rm NMR}\!\approx\!1$ above 100 K implying uniform relaxation, and a decreasing $\beta_{\rm NMR}< 1$ below 100 K signaling inhomogeneous dynamics.
At 0.67 T, both $\beta_{\rm NMR}$ and $1/T_1$ exhibit a broad $\sim$1 K hump (seen in Figure \ref{fig5}c and \ref{fig5}d), consistent with the broad anomaly in $C_p$ and the increases of $\lambda_{\mu{\rm SR}}$ and $\beta_{\mu{\rm SR}}$ near 1 K. 
Moreover, for 0.67 T the spectral intensity multiplied by temperature $I\times T$ drops below $\sim$1 K without a shape change (seen in Figure \ref{fig5}a), pointing to an inhomogeneous distribution of static internal fields as also seen by $\mu$SR.
The intensity loss likely reflects a subset of $^{23}$Na sites experiencing static fields that broaden or shift the resonance, leaving the low-$T$ signal from $^{23}$Na near Cu moments that remain dynamic.
In contrast, at 3 T the $I\times T$ is nearly temperature-independent because the orphan spins that were responsible for the static-field broadening at 0.67 T are now mostly polarized, as can be inferred from the orphan-spin Brillouin function in Figure \ref{fig2}d. Correspondingly, the $\sim$1 K anomaly in $\beta_{\rm NMR}$ is suppressed at 3 T, and the $1/T_1$ hump is smoothed and shifts to higher temperature.

\subsection{Scaling behaviour}

We now turn to the scaling behaviour and data collapse of measured physical quantities $\chi(T)$, $C_{\rm mag}/T$ and $^{23}(1/T_1T)$ in $T/(\mu_0 H)$, as well as $M(H)$ in $(\mu_0 H)/T$ and $P(t)$ in $t/(\mu_0H)^{\eta_{\rm \mu SR}}$. This behaviour provides the smoking gun evidence for the RS ground state in NaCdCu$_2$F$_7$, in agreement with the scaling theory for $S=\frac{1}{2}$ frustrated quantum magnets with quenched disorder \cite{Kimchi2018b, Kimchi2018,Dasgupta1980}.

\begin{figure}
    \centering
    \includegraphics[width=1\linewidth]{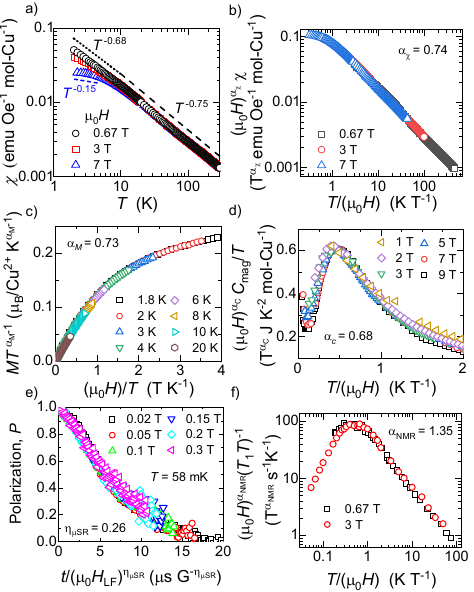}
    \caption{(a) Power-law scaling of DC magnetic susceptibility $\chi(T)\sim T^{-\alpha_\chi}$ at 0.67 T, 3 T and 7 T fields, with (b) a data collapse of $(\mu_0 H)^{\alpha_\chi}\chi$ vs. $T/(\mu_0 H)$ for $\alpha_\chi=0.74$. Data collapse of (c) isothermal magnetization scaled by temperature $MT^{\alpha_M-1}$ vs. $(\mu_0H)/T$ for $\alpha_M=0.73$, (d) magnetic specific heat scaled by field and temperature $(\mu_0H)^{\alpha_C}C_{\rm mag}/T$ vs. $T/(\mu_0H)$ for $\alpha_C=0.68$, (e) muon polarization $P(t)$ as a function of $t/(\mu_0H_{\rm LF})^{\eta_{\rm \mu SR}}$ for $\eta_{\rm \mu SR} = 1-\alpha_{\rm \mu SR} = 0.26$, and (f) the spin-lattice relaxation rate scaled by field and temperature $(\mu_0H)^{\alpha_{\rm NMR}}(T_1T)^{-1}$ vs. $T/(\mu_0H)$ for $\alpha_{\rm NMR}=1.35$.}
    \label{fig6}
\end{figure}

In Figure \ref{fig6}a, we show the temperature-dependent DC susceptibility in multiple fields, on a log-log scale. $\chi(T)$ exhibits two distinct power-law regimes as $\chi(T) \sim T^{-\alpha_\chi}$: a field-independent regime above 10 K with $\alpha_\chi = 0.75(1)$, as well as a field-dependent regime below 10 K with varying $\alpha_\chi$ = 0.15(1) -- 0.68(1), decreasing with applied field as $\chi(T)$ flattens upon polarization of the orphan spins. In addition, the susceptibility data in different fields collapse onto a single curve after scaling by field $(\mu_0 H)^{\alpha_\chi} \chi(T) $ as a function of $T/(\mu_0 H)$, with $\alpha_\chi = 0.74$, see Figure \ref{fig6}b.
Figure \ref{fig6}c shows the data collapse of isothermal magnetization at temperatures between 1.8 and 20 K, onto a single curve after scaling by temperature $M T^{\alpha_M -1}$ as a function of $\mu_0 H/T$ \cite{Kimchi2018}. This displays a perfect data collapse for $\alpha_M = 0.73$, in agreement with the fitted exponent in the power-law term in Equation \ref{eq1}. Furthermore, in Figure \ref{fig6}d, we observe a data collapse of magnetic specific heat scaled by field $(\mu_0 H)^{\alpha_C}  C_{\rm mag}/T$ as a function of $T/(\mu_0 H)$ for $\alpha_C = 0.68$. Strikingly, the power-law scaling also applies to the muon relaxation rate $\lambda_{\rm LF} \sim (\mu_0 H_{LF})^{-\eta_{\rm \mu SR}}$ acquired from fitting the field-dependent spectra between 200 and 3000 G by a stretched exponential function (see Figure S9 and S10 in SM \cite{Supplementary}). Here, $\eta_{\rm \mu SR} = 1-\alpha_{\rm \mu SR}$ = 0.26(2), giving $\alpha_{\rm \mu SR} = 0.74(2)$. This leads to a data collapse of the field-dependent muon polarization spectra $P(t)=(A_{\rm LF}(t)-A_{\rm base})/A_0$ when plotted as a function of $t/(\mu_0 H)^{\eta_{\rm \mu SR}}$, see Figure \ref{fig6}e. The observed power-law behaviour is a result of a power-law decaying spin-spin correlation function $S(t) \sim t^{-(1-\alpha)}$ governed by a power-law spectral density $S(\omega) \sim \omega^{-\alpha}$ \cite{Pratt2006}, which have been reported in several other QSL candidates with spin chain, pyrochlore, kagome, and triangular lattices \cite{Pratt2006,Lee2020, Keren2004,Kermarrec2014,Li2016}. The power-law scaling is also evident in the $^{23}$Na NMR as $1/T_1 \sim T^{\eta_{\rm NMR}}$, exhibiting $\eta_{\rm NMR}$ = 1.34(7) at 0.67 T below 400 mK, and $\eta_{\rm NMR}$ = 2.41(4) at 3 T below 1 K (see Figure \ref{fig5}c), consistent with other RS and QSL candidates \cite{Lee2023,Shimizu2006}. This leads to a data collapse of $^{23}(1/T_1T)$ scaled by field $(\mu_0 H)^{\alpha_{\rm NMR}}(1/T_1T)$ as a function of $T/(\mu_0 H)$ for $\alpha_{\rm NMR} = 1.35$, see Figure \ref{fig6}f. We note that the exponent $\alpha_{\rm NMR} = 1.35$ is approximately double in value compared to $\alpha_\chi$, $\alpha_M$, $\alpha_C$ or $\alpha_{\mu \rm SR}$. While the bulk probes $\chi(T)$, $M(H)$ or $C_{\rm mag}/T$ reflect the static density of low-energy excitations, the quantity $(1/T_1T) \sim \sum\limits_\textbf{q} |A_\textbf{q}|^2 \left( \chi''(\textbf{q},\omega_0)/\omega_0 \right) _{\omega_0 \rightarrow 0}$ is proportional to the momentum $\textbf{q}$-averaged imaginary part of the local spin susceptibility, reflecting the low-frequency spectral weight of spin fluctuations. Although LF-$\mu$SR also probes the low-frequency limit of the dynamical susceptibility as $\lambda_{\rm LF} \sim \sum\limits_\textbf{q} |A^{\mu}_\textbf{q}|^2 \left( \chi''(\textbf{q},\omega_{\mu})/\omega_{\mu} \right)_{\omega_{\mu}\rightarrow 0}$, the $\alpha_{\mu \rm SR} = 0.74(2)$ exponent agrees more with the bulk-probe exponents rather than $\alpha_{\rm NMR}$. We attribute this to the different frequency windows in which these probes operate: LF-$\mu$SR is sensitive to fluctuations in the $\sim 10^{5}$--$10^{7}$ Hz range set by the muon gyromagnetic ratio ($\omega_\mu = \gamma_\mu H_{\rm LF}$), whereas $^{23}$Na NMR probes dynamics at the Larmor frequency at $\sim 10^{7}$--$10^{8}$ Hz ($\omega_0 = {}^{23}\!\gamma_{\mathrm{Na}}
 H_{\rm LF}$). In a random singlet system with a broad distribution of excitation energies, the ultra-low-frequency window accessed by LF-$\mu$SR emphasizes the same low-energy excitations that dominate bulk thermodynamics, while NMR probes higher-frequency dynamics, resulting in different scaling exponents.

\subsection{Conclusion}

In summary, using a plethora of complementary bulk and local-probe techniques, we have shown that the frustrated, magnetic-bond-disordered $S=\frac{1}{2}$ Heisenberg pyrochlore antiferromagnet NaCdCu$_2$F$_7$ does not exhibit long-range magnetic order or global spin freezing down to 58 mK. Instead, the observed power-law scaling behaviour and data collapse across multiple experimental probes are consistent with a random singlet ground state characterized by a broad distribution of exchange couplings, noted by the coexistence of singlet dimers and weakly coupled orphan spins. This behavior originates from intrinsic Na$^+$/Cd$^{2+}$ disorder on the pyrochlore $A$ site, which modifies the Cu$^{2+}$-F$^-$-Cu$^{2+}$ superexchange pathways on the magnetic, pyrochlore $B$-site sublattice. The absence of spin-glass freezing contrasts with previously studied Na$A''B_2$F$_7$ pyrochlore fluorides \cite{Kancko2023,Kancko2025,Krizan2014,Krizan2015Co,Krizan2015Ni,Sanders2017}, where comparable magnetic bond disorder leads to glassy ground states, suggesting that strong $S=\tfrac{1}{2}$ quantum fluctuations play an important role in suppressing spin freezing in NaCdCu$_2$F$_7$ and stabilizing a quantum-disordered state.

The present study establishes NaCdCu$_2$F$_7$ as a well-characterized example of disorder-driven quantum magnetism, but several important questions remain. All measurements reported here were performed on polycrystalline samples, and the growth of high-quality single crystals would enable momentum-resolved and anisotropic measurements. Thermal transport experiments would be particularly valuable for probing low-energy excitations and further distinguishing random singlet behavior from other quantum-disordered states. Neutron scattering studies would provide direct access to the dynamic spin structure factor; however, such experiments are presently hindered by the strong neutron absorption of Cd, motivating isotopic enrichment or the investigation of chemically related pyrochlore fluorides such as NaCaCu$_2$F$_7$ or NaSrCu$_2$F$_7$.

Finally, recent theoretical work has shown that entanglement-sensitive quantities such as quantum Fisher information (QFI), extracted from experimentally accessible correlation functions \cite{Shimokawa2025}, may distinguish random singlet states from quantum spin liquids even when thermodynamic signatures are similar. Applying such approaches to NaCdCu$_2$F$_7$ and related materials represents an interesting direction for future experimental and theoretical studies.

\begin{acknowledgments}
We acknowledge funding from Charles University in Prague, within Grant Agency Univerzita Karlova program (GAUK 48924). Crystal growth, structural analysis and magnetic properties measurements were carried out in the MGML (http://mgml.eu/), which is supported within the program of Czech Research Infrastructures (project no. LM2023065). This work was supported by the Ministry of Education, Youth and Sports of the Czech Republic through the INTER-EXCELLENCE II program (LUABA24056), by the Grant Agency of the Czech Republic (26-23051S), as well as by the HLD at HZDR, member of the European Magnetic Field Laboratory (EMFL) and by EPSRC (UK) via its membership to the EMFL (grant no. EP/N01085X/1). Muon spectroscopy measurements were supported by the ISIS neutron and muon source \cite{Colman2024}.
H. S. and Y. T. acknowledge support from Japan Society for the Promotion of Science (JSPS) through KAKENHI (JP24KK0062 and JP23K25829).
Work at the Japan Atomic Energy Agency (JAEA) was partially supported by the JAEA REIMEI Research Program.
The authors additionally thank Gael Bastien for insightful discussions, and Martin Misek for performing complementary magnetization measurements.

\end{acknowledgments}

\bibliography{bibliography}

@article{Balents2010,
author = {Balents, Leon},
doi = {10.1038/nature08917},
issn = {0028-0836},
journal = {Nature},
month = {mar},
number = {7286},
pages = {199--208},
title = {{Spin liquids in frustrated magnets}},
url = {https://www.nature.com/articles/nature08917},
volume = {464},
year = {2010}
}

@article{Savary2017,
author = {Savary, Lucile and Balents, Leon},
doi = {10.1088/0034-4885/80/1/016502},
issn = {0034-4885},
journal = {Reports on Progress in Physics},
month = {jan},
number = {1},
pages = {016502},
pmid = {27823986},
publisher = {IOP Publishing},
title = {{Quantum spin liquids: a review}},
url = {https://iopscience.iop.org/article/10.1088/0034-4885/80/1/016502},
volume = {80},
year = {2017}
}

@article{Lancaster2023,
archivePrefix = {arXiv},
arxivId = {2310.19577},
author = {Lancaster, Tom},
doi = {10.1080/00107514.2023.2284522},
eprint = {2310.19577},
issn = {0010-7514},
journal = {Contemporary Physics},
month = {apr},
number = {2},
pages = {127--146},
publisher = {Taylor & Francis},
title = {{Quantum spin liquids}},
url = {https://doi.org/10.1080/00107514.2023.2284522 https://www.tandfonline.com/doi/full/10.1080/00107514.2023.2284522},
volume = {64},
year = {2023}
}

@article{A.Yu.Kitaev2002,
author = {Kitaev, A.Yu.},
doi = {10.1016/S0003-4916(02)00018-0},
issn = {00034916},
journal = {Annals of Physics},
month = {jan},
number = {1},
pages = {2--30},
title = {{Fault-tolerant quantum computation by anyons}},
url = {https://linkinghub.elsevier.com/retrieve/pii/S0003491602000180},
volume = {303},
year = {2003}
}

@article{Yang2018,
archivePrefix = {arXiv},
arxivId = {1801.05698},
author = {Yang, Fan and Henriet, Lo{\"{i}}c and Soret, Ariane and {Le Hur}, Karyn},
doi = {10.1103/PhysRevB.98.035431},
eprint = {1801.05698},
issn = {24699969},
journal = {Physical Review B},
number = {3},
publisher = {American Physical Society},
title = {{Engineering quantum spin liquids and many-body Majorana states with a driven superconducting box circuit}},
volume = {98},
year = {2018}
}

@article{Nayak2008,
archivePrefix = {arXiv},
arxivId = {0707.1889},
author = {Nayak, Chetan and Simon, Steven H. and Stern, Ady and Freedman, Michael and {Das Sarma}, Sankar},
doi = {10.1103/RevModPhys.80.1083},
eprint = {0707.1889},
issn = {0034-6861},
journal = {Reviews of Modern Physics},
month = {sep},
number = {3},
pages = {1083--1159},
title = {{Non-Abelian anyons and topological quantum computation}},
url = {https://link.aps.org/doi/10.1103/RevModPhys.80.1083},
volume = {80},
year = {2008}
}

@article{Kitagawa2018,
author = {Kitagawa, K. and Takayama, T. and Matsumoto, Y. and Kato, A. and Takano, R. and Kishimoto, Y. and Bette, S. and Dinnebier, R. and Jackeli, G. and Takagi, H.},
doi = {10.1038/nature25482},
issn = {0028-0836},
journal = {Nature},
month = {feb},
number = {7692},
pages = {341--345},
pmid = {29446382},
publisher = {Nature Publishing Group},
title = {{A spin--orbital-entangled quantum liquid on a honeycomb lattice.}},
url = {https://www.nature.com/articles/nature25482},
volume = {554},
year = {2018}
}

@article{Zhou2017,
archivePrefix = {arXiv},
arxivId = {1607.03228},
author = {Zhou, Yi and Kanoda, Kazushi and Ng, Tai-Kai},
doi = {10.1103/RevModPhys.89.025003},
eprint = {1607.03228},
issn = {0034-6861},
journal = {Reviews of Modern Physics},
month = {apr},
number = {2},
pages = {025003},
title = {{Quantum spin liquid states}},
url = {http://link.aps.org/doi/10.1103/RevModPhys.89.025003},
volume = {89},
year = {2017}
}

@article{Harris1991,
author = {Harris, A B and Berlinsky, A J and Bruder, C},
doi = {10.1063/1.348098},
issn = {0021-8979},
journal = {Journal of Applied Physics},
month = {apr},
number = {8},
pages = {5200--5202},
title = {{Ordering by quantum fluctuations in a strongly frustrated Heisenberg antiferromagnet}},
url = {https://pubs.aip.org/jap/article/69/8/5200/1066604/Ordering-by-quantum-fluctuations-in-a-strongly},
volume = {69},
year = {1991}
}

@article{Canals2000,
author = {Canals, Benjamin and Lacroix, Claudine},
doi = {10.1103/PhysRevB.61.1149},
issn = {0163-1829},
journal = {Physical Review B},
month = {jan},
number = {2},
pages = {1149--1159},
title = {{Quantum spin liquid: The Heisenberg antiferromagnet on the three-dimensional pyrochlore lattice}},
url = {https://link.aps.org/doi/10.1103/PhysRevB.61.1149},
volume = {61},
year = {2000}
}

@article{Kimchi2018,
archivePrefix = {arXiv},
arxivId = {1803.00013},
author = {Kimchi, Itamar and Sheckelton, John P. and McQueen, Tyrel M. and Lee, Patrick A.},
doi = {10.1038/s41467-018-06800-2},
eprint = {1803.00013},
issn = {2041-1723},
journal = {Nature Communications},
month = {oct},
number = {1},
pages = {4367},
pmid = {30349043},
publisher = {Springer US},
title = {{Scaling and data collapse from local moments in frustrated disordered quantum spin systems}},
url = {http://dx.doi.org/10.1038/s41467-018-06800-2 https://www.nature.com/articles/s41467-018-06800-2},
volume = {9},
year = {2018}
}

@article{Schafer2020,
archivePrefix = {arXiv},
arxivId = {2003.04898},
author = {Sch{\"{a}}fer, Robin and Hagym{\'{a}}si, Imre and Moessner, Roderich and Luitz, David J.},
doi = {10.1103/PhysRevB.102.054408},
eprint = {2003.04898},
issn = {2469-9950},
journal = {Physical Review B},
month = {aug},
number = {5},
pages = {054408},
title = {{Pyrochlore $S=\frac{1}{2}$ Heisenberg antiferromagnet at finite temperature}},
url = {https://link.aps.org/doi/10.1103/PhysRevB.102.054408},
volume = {102},
year = {2020}
}

@article{Kim2008,
author = {Kim, Jung Hoon and Han, Jung Hoon},
doi = {10.1103/PhysRevB.78.180410},
issn = {10980121},
journal = {Physical Review B - Condensed Matter and Materials Physics},
number = {18},
pages = {1--4},
title = {{Chiral spin states in the pyrochlore Heisenberg magnet: Fermionic mean-field theory and variational Monte Carlo calculations}},
volume = {78},
year = {2008}
}

@article{Tsunetsugu2001,
author = {Tsunetsugu, Hirokazu},
doi = {10.1103/PhysRevB.65.024415},
issn = {0163-1829},
journal = {Physical Review B},
month = {dec},
number = {2},
pages = {024415},
title = {{Spin-singlet order in a pyrochlore antiferromagnet}},
url = {https://link.aps.org/doi/10.1103/PhysRevB.65.024415},
volume = {65},
year = {2001}
}

@article{Astrakhantsev2021,
archivePrefix = {arXiv},
arxivId = {2101.08787},
author = {Astrakhantsev, Nikita and Westerhout, Tom and Tiwari, Apoorv and Choo, Kenny and Chen, Ao and Fischer, Mark H. and Carleo, Giuseppe and Neupert, Titus},
doi = {10.1103/PhysRevX.11.041021},
eprint = {2101.08787},
issn = {2160-3308},
journal = {Physical Review X},
month = {oct},
number = {4},
pages = {041021},
publisher = {American Physical Society},
title = {{Broken-Symmetry Ground States of the Heisenberg Model on the Pyrochlore Lattice}},
url = {https://doi.org/10.1103/PhysRevX.11.041021 https://link.aps.org/doi/10.1103/PhysRevX.11.041021},
volume = {11},
year = {2021}
}

@article{Chandra2018,
archivePrefix = {arXiv},
arxivId = {1710.11316},
author = {Chandra, V. Ravi and Sahoo, Jyotisman},
doi = {10.1103/PhysRevB.97.144407},
eprint = {1710.11316},
issn = {2469-9950},
journal = {Physical Review B},
month = {apr},
number = {14},
pages = {144407},
publisher = {American Physical Society},
title = {{Spin-$\frac{1}{2}$ Heisenberg antiferromagnet on the pyrochlore lattice: An exact diagonalization study}},
url = {https://link.aps.org/doi/10.1103/PhysRevB.97.144407},
volume = {97},
year = {2018}
}

@article{Schneider2022,
archivePrefix = {arXiv},
arxivId = {2107.13574},
author = {Schneider, Benedikt and Halimeh, Jad C. and Punk, Matthias},
doi = {10.1103/PhysRevB.105.125122},
eprint = {2107.13574},
issn = {24699969},
journal = {Physical Review B},
number = {12},
pages = {1--20},
publisher = {American Physical Society},
title = {{Projective symmetry group classification of chiral $Z_{2}$ spin liquids on the pyrochlore lattice: Application to the spin-$\frac{1}{2}$ $XXZ$ Heisenberg model}},
volume = {105},
year = {2022}
}

@article{Hagymasi2021,
archivePrefix = {arXiv},
arxivId = {2010.03563},
author = {Hagym{\'{a}}si, Imre and Sch{\"{a}}fer, Robin and Moessner, Roderich and Luitz, David J.},
doi = {10.1103/PhysRevLett.126.117204},
eprint = {2010.03563},
issn = {0031-9007},
journal = {Physical Review Letters},
month = {mar},
number = {11},
pages = {117204},
pmid = {33798350},
publisher = {American Physical Society},
title = {{Possible Inversion Symmetry Breaking in the $S=\frac{1}{2}$ Pyrochlore Heisenberg Magnet}},
url = {https://doi.org/10.1103/PhysRevLett.126.117204 https://link.aps.org/doi/10.1103/PhysRevLett.126.117204},
volume = {126},
year = {2021}
}

@article{Gardner1999,
author = {Gardner, J. S. and Dunsiger, S. R. and Gaulin, B. D. and Gingras, M. J. P. and Greedan, J. E. and Kiefl, R. F. and Lumsden, M. D. and MacFarlane, W. A. and Raju, N. P. and Sonier, J. E. and Swainson, I. and Tun, Z.},
doi = {10.1103/PhysRevLett.82.1012},
issn = {0031-9007},
journal = {Physical Review Letters},
month = {feb},
number = {5},
pages = {1012--1015},
title = {{Cooperative Paramagnetism in the Geometrically Frustrated Pyrochlore Antiferromagnet ${\mathrm{Tb}}_{2}{\mathrm{Ti}}_{2}{\mathrm{O}}_{7}$}},
url = {https://link.aps.org/doi/10.1103/PhysRevLett.82.1012},
volume = {82},
year = {1999}
}

@article{Nakatsuji2006,
author = {Nakatsuji, S. and Machida, Y. and Maeno, Y. and Tayama, T. and Sakakibara, T. and van Duijn, J. and Balicas, L. and Millican, J. N. and Macaluso, R. T. and Chan, Julia Y.},
doi = {10.1103/PhysRevLett.96.087204},
issn = {0031-9007},
journal = {Physical Review Letters},
month = {mar},
number = {8},
pages = {087204},
title = {{Metallic Spin-Liquid Behavior of the Geometrically Frustrated Kondo Lattice ${\mathrm{Pr}}_{2}{\mathrm{Ir}}_{2}{\mathrm{O}}_{7}$}},
url = {https://link.aps.org/doi/10.1103/PhysRevLett.96.087204},
volume = {96},
year = {2006}
}

@article{Sibille2015,
author = {Sibille, Romain and Lhotel, Elsa and Pomjakushin, Vladimir and Baines, Chris and Fennell, Tom and Kenzelmann, Michel},
doi = {10.1103/PhysRevLett.115.097202},
issn = {10797114},
journal = {Physical Review Letters},
number = {9},
pages = {1--5},
title = {{Candidate Quantum Spin Liquid in the ${\mathrm{Ce}}^{3+}$ Pyrochlore Stannate ${\mathrm{Ce}}_{2}{\mathrm{Sn}}_{2}{\mathrm{O}}_{7}$}},
volume = {115},
year = {2015}
}

@article{Uematsu2019,
archivePrefix = {arXiv},
arxivId = {1901.10291},
author = {Uematsu, Kazuki and Kawamura, Hikaru},
doi = {10.1103/PhysRevLett.123.087201},
eprint = {1901.10291},
issn = {0031-9007},
journal = {Physical Review Letters},
month = {aug},
number = {8},
pages = {087201},
pmid = {31491226},
publisher = {American Physical Society},
title = {{Randomness-Induced Quantum Spin Liquid Behavior in the $S=\frac{1}{2}$ Random-Bond Heisenberg Antiferromagnet on the Pyrochlore Lattice}},
url = {https://doi.org/10.1103/PhysRevLett.123.087201 https://link.aps.org/doi/10.1103/PhysRevLett.123.087201},
volume = {123},
year = {2019}
}

@article{Isoda1998,
author = {Isoda, Makoto and Mori, Shigeyoshi},
doi = {10.1143/JPSJ.67.4022},
issn = {00319015},
journal = {Journal of the Physical Society of Japan},
number = {12},
pages = {4022--4025},
title = {{Valence-Bond Crystal and Anisotropic Excitation Spectrum on 3-Dimensionally Frustrated Pyrochlore}},
volume = {67},
year = {1998}
}

@article{Clark2014,
archivePrefix = {arXiv},
arxivId = {1405.3172},
author = {Clark, L. and Nilsen, G. J. and Kermarrec, E. and Ehlers, G. and Knight, K. S. and Harrison, A. and Attfield, J. P. and Gaulin, B. D.},
doi = {10.1103/PhysRevLett.113.117201},
eprint = {1405.3172},
issn = {0031-9007},
journal = {Physical Review Letters},
month = {sep},
number = {11},
pages = {117201},
title = {{From Spin Glass to Quantum Spin Liquid Ground States in Molybdate Pyrochlores}},
url = {https://link.aps.org/doi/10.1103/PhysRevLett.113.117201},
volume = {113},
year = {2014}
}

@article{Kawabata2007,
author = {Kawabata, Shohei and Yasui, Yukio and Kobayashi, Yoshiaki and Sato, Masatoshi},
doi = {10.1143/JPSJ.76.084705},
issn = {00319015},
journal = {Journal of the Physical Society of Japan},
number = {8},
pages = {1--5},
title = {{Magnetic Behavior of Spin-$\frac{1}{2}$ Frustrated System Hg$_2$Cu$_2$F$_6$S}},
volume = {76},
year = {2007}
}

@article{Volkov2020,
archivePrefix = {arXiv},
arxivId = {2001.06013},
author = {Volkov, Pavel A. and Won, Choong-Jae and Gorbunov, D. I. and Kim, Jaewook and Ye, Mai and Kim, Heung-Sik and Pixley, J. H. and Cheong, Sang-Wook and Blumberg, G.},
doi = {10.1103/PhysRevB.101.020406},
eprint = {2001.06013},
issn = {2469-9950},
journal = {Physical Review B},
month = {jan},
number = {2},
pages = {020406},
publisher = {American Physical Society},
title = {{Random singlet state in ${\mathrm{Ba}}_{5}{\mathrm{CuIr}}_{3}{\mathrm{O}}_{12}$ single crystals}},
url = {https://doi.org/10.1103/PhysRevB.101.020406 https://link.aps.org/doi/10.1103/PhysRevB.101.020406},
volume = {101},
year = {2020}
}

@article{Kundu2020,
archivePrefix = {arXiv},
arxivId = {2009.05137},
author = {Kundu, S. and Hossain, Akmal and S., Pranava Keerthi and Das, Ranjan and Baenitz, M. and Baker, Peter J. and Orain, Jean-Christophe and Joshi, D. C. and Mathieu, Roland and Mahadevan, Priya and Pujari, Sumiran and Bhattacharjee, Subhro and Mahajan, A. V. and Sarma, D. D.},
doi = {10.1103/PhysRevLett.125.117206},
eprint = {2009.05137},
issn = {0031-9007},
journal = {Physical Review Letters},
month = {sep},
number = {11},
pages = {117206},
pmid = {32975979},
publisher = {American Physical Society},
title = {{Signatures of a $S=\frac{1}{2}$ Cooperative Paramagnet in the Diluted Triangular Lattice of ${\mathrm{Y}}_{2}{\mathrm{CuTiO}}_{6}$}},
url = {https://doi.org/10.1103/PhysRevLett.125.117206 https://link.aps.org/doi/10.1103/PhysRevLett.125.117206},
volume = {125},
year = {2020}
}

@article{Lee2023,
author = {Lee, Chanhyeon and Lee, Suheon and Choi, Youngsu and Wang, C. and Luetkens, H. and Shiroka, T. and Jang, Zeehoon and Yoon, Young-Gui and Choi, Kwang-Yong},
doi = {10.1103/PhysRevB.107.014424},
issn = {2469-9950},
journal = {Physical Review B},
month = {jan},
number = {1},
pages = {014424},
publisher = {American Physical Society},
title = {{Coexistence of random singlets and disordered Kitaev spin liquid in ${\mathrm{H}}_{3}{\mathrm{LiIr}}_{2}{\mathrm{O}}_{6}$}},
url = {https://link.aps.org/doi/10.1103/PhysRevB.107.014424},
volume = {107},
year = {2023}
}

@article{Sheckelton2012,
author = {Sheckelton, J. P. and Neilson, J. R. and Soltan, D. G. and McQueen, T. M.},
doi = {10.1038/nmat3329},
issn = {1476-1122},
journal = {Nature Materials},
month = {jun},
number = {6},
pages = {493--496},
publisher = {Nature Publishing Group},
title = {{Possible valence-bond condensation in the frustrated cluster magnet ${\mathrm{LiZn}}_{2}{\mathrm{Mo}}_{3}{\mathrm{O}}_{8}$}},
url = {http://dx.doi.org/10.1038/nmat3329 https://www.nature.com/articles/nmat3329},
volume = {11},
year = {2012}
}

@article{Murayama2022,
archivePrefix = {arXiv},
arxivId = {2106.07223},
author = {Murayama, H. and Tominaga, T. and Asaba, T. and Silva, A. de Oliveira and Sato, Y. and Suzuki, H. and Ukai, Y. and Suetsugu, S. and Kasahara, Y. and Okuma, R. and Kimchi, I. and Matsuda, Y.},
doi = {10.1103/PhysRevB.106.174406},
eprint = {2106.07223},
issn = {24699969},
journal = {Physical Review B},
number = {17},
pages = {1--8},
publisher = {American Physical Society},
title = {{Universal scaling of specific heat in the $S = \frac{1}{2}$ quantum kagome antiferromagnet herbertsmithite}},
volume = {106},
year = {2022}
}

@article{Pal2020,
archivePrefix = {arXiv},
arxivId = {1906.10914},
author = {Pal, Sudip and Kumar, Kranti and Sharma, Rohit and Banerjee, A. and Roy, S. B. and Park, Je-Geun and Nigam, A. K. and Cheong, Sang-Wook},
doi = {10.1088/1361-648X/ab48be},
eprint = {1906.10914},
issn = {0953-8984},
journal = {Journal of Physics: Condensed Matter},
month = {jan},
number = {3},
pages = {035601},
pmid = {31561241},
publisher = {IOP Publishing},
title = {{Possible glass-like random singlet magnetic state in ${\mathrm{1T}-\mathrm{TaS}}_{2}$ }},
url = {https://iopscience.iop.org/article/10.1088/1361-648X/ab48be},
volume = {32},
year = {2020}
}

@article{Dasgupta1980,
author = {Dasgupta, Chandan and Ma, Shang-Keng},
doi = {10.1103/PhysRevB.22.1305},
issn = {0163-1829},
journal = {Physical Review B},
month = {aug},
number = {3},
pages = {1305--1319},
title = {{Low-temperature properties of the random Heisenberg antiferromagnetic chain}},
url = {https://link.aps.org/doi/10.1103/PhysRevB.22.1305},
volume = {22},
year = {1980}
}

@article{Fisher1994,
author = {Fisher, Daniel S.},
doi = {10.1103/PhysRevB.50.3799},
issn = {0163-1829},
journal = {Physical Review B},
month = {aug},
number = {6},
pages = {3799--3821},
title = {{Random antiferromagnetic quantum spin chains}},
url = {https://link.aps.org/doi/10.1103/PhysRevB.50.3799},
volume = {50},
year = {1994}
}

@article{Bhatt1982,
author = {Bhatt, R. N. and Lee, P. A.},
doi = {10.1103/PhysRevLett.48.344},
issn = {0031-9007},
journal = {Physical Review Letters},
month = {feb},
number = {5},
pages = {344--347},
title = {{Scaling Studies of Highly Disordered spin-$\frac{1}{2}$ Antiferromagnetic Systems}},
url = {https://link.aps.org/doi/10.1103/PhysRevLett.48.344},
volume = {48},
year = {1982}
}

@article{Kancko2023,
author = {Kancko, A. and Giester, G. and Colman, R. H.},
doi = {10.1088/1402-4896/acdeb7},
issn = {0031-8949},
journal = {Physica Scripta},
month = {jul},
number = {7},
pages = {075947},
publisher = {IOP Publishing},
title = {{Structural and spin-glass properties of single crystal ${J}_{\mathrm{eff}} = \frac{1}{2}$ pyrochlore antiferromagnet ${\text{NaCdCo}}_{2}{\mathrm{F}}_{7}$: correlating $T_f$ with magnetic-bond-disorder}},
url = {https://iopscience.iop.org/article/10.1088/1402-4896/acdeb7},
volume = {98},
year = {2023}
}

@article{Krizan2014,
author = {Krizan, J. W. and Cava, R. J.},
doi = {10.1103/PhysRevB.89.214401},
issn = {1098-0121},
journal = {Physical Review B},
month = {jun},
number = {21},
pages = {214401},
title = {{${\text{NaCaCo}}_{2}{\mathrm{F}}_{7}$: A single-crystal high-temperature pyrochlore antiferromagnet}},
url = {https://link.aps.org/doi/10.1103/PhysRevB.89.214401},
volume = {89},
year = {2014}
}

@article{Krizan2015Co,
author = {Krizan, J. W. and Cava, R. J.},
doi = {10.1088/0953-8984/27/29/296002},
issn = {0953-8984},
journal = {Journal of Physics: Condensed Matter},
month = {jul},
number = {29},
pages = {296002},
publisher = {IOP Publishing},
title = {{NaSrCo$_{2}$F$_{7}$, a Co$^{{2+}}$ pyrochlore antiferromagnet}},
url = {http://dx.doi.org/10.1088/0953-8984/27/29/296002 https://iopscience.iop.org/article/10.1088/0953-8984/27/29/296002},
volume = {27},
year = {2015}
}

@article{Krizan2015Ni,
archivePrefix = {arXiv},
arxivId = {1504.07708},
author = {Krizan, J. W. and Cava, R. J.},
doi = {10.1103/PhysRevB.92.014406},
eprint = {1504.07708},
issn = {1098-0121},
journal = {Physical Review B},
month = {jul},
number = {1},
pages = {014406},
title = {{${\text{NaCaNi}}_{2}{\mathrm{F}}_{7}$: A frustrated high temperature pyrochlore antiferromagnet with $S=1$ ${\mathrm{Ni}}^{2+}$}},
url = {https://link.aps.org/doi/10.1103/PhysRevB.92.014406},
volume = {92},
year = {2015}
}

@article{Oliveira2004,
author = {Oliveira, E. A. and Guedes, I. and Ayala, A. P. and Gesland, J. Y. and Ellena, J. and Moreira, R. L. and Grimsditch, M.},
doi = {10.1016/j.jssc.2004.04.055},
issn = {00224596},
journal = {Journal of Solid State Chemistry},
number = {8},
pages = {2943--2950},
title = {{Crystal structure and vibrational spectrum of the ${\text{NaCaMg}}_{2}{\mathrm{F}}_{7}$ pyrochlore}},
volume = {177},
year = {2004}
}

@article{Kubel2001,
author = {Kubel, Frank and Dundjerski, Barbara},
doi = {10.1002/1521-3749(200107)627:7<1589::aid-zaac1589>3.0.co;2-2},
issn = {00442313},
journal = {Zeitschrift fur Anorganische und Allgemeine Chemie},
number = {7},
pages = {1589--1592},
title = {{Synthesis and crystal structure analysis of 
${\text{NaSrMg}}_{2}{\mathrm{F}}_{7}$, a fully fluorinated compound of the pyrochlore family}},
volume = {627},
year = {2001}
}

@article{Grzechnik2007,
author = {Grzechnik, Andrzej and Kaindl, Reinhard and Friese, Karen},
doi = {10.1016/j.jpcs.2006.11.029},
issn = {00223697},
journal = {Journal of Physics and Chemistry of Solids},
number = {3},
pages = {382--388},
title = {{Temperature dependent study of the crystal structure of ${\text{NaCdZn}}_{2}{\mathrm{F}}_{7}$ pyrochlore}},
volume = {68},
year = {2007}
}

@article{Sanders2017,
author = {Sanders, M. B. and Krizan, J. W. and Plumb, K. W. and McQueen, T. M. and Cava, R. J.},
doi = {10.1088/1361-648X/29/4/045801},
issn = {0953-8984},
journal = {Journal of Physics: Condensed Matter},
month = {feb},
number = {4},
pages = {045801},
pmid = {27875333},
publisher = {IOP Publishing},
title = {{${\text{NaSrMn}}_{2}{\mathrm{F}}_{7}$, ${\text{NaCaFe}}_{2}{\mathrm{F}}_{7}$, and ${\text{NaSrFe}}_{2}{\mathrm{F}}_{7}$: novel single crystal pyrochlore antiferromagnets}},
url = {https://iopscience.iop.org/article/10.1088/1361-648X/29/4/045801},
volume = {29},
year = {2017}
}

@article{Sarkar2017,
archivePrefix = {arXiv},
arxivId = {1604.00814},
author = {Sarkar, R. and Krizan, J. W. and Br{\"{u}}ckner, F. and Andrade, E. C. and Rachel, S. and Vojta, M. and Cava, R. J. and Klauss, H. H.},
doi = {10.1103/PhysRevB.96.235117},
eprint = {1604.00814},
issn = {24699969},
journal = {Physical Review B},
number = {23},
pages = {1--10},
title = {{Spin freezing in the disordered pyrochlore magnet ${\text{NaCaCo}}_{2}{\mathrm{F}}_{7}$: NMR studies and Monte Carlo simulations}},
volume = {96},
year = {2017}
}

@article{Hansler1970,
author = {H{\"{a}}nsler, R. and R{\"{u}}dorff, W.},
doi = {10.1515/znb-1970-1121},
issn = {1865-7117},
journal = {Zeitschrift f{\"{u}}r Naturforschung B},
month = {nov},
number = {11},
pages = {1306--1307},
title = {{${\text{A}}_{2}{\mathrm{B}}_{7}{\mathrm{F}}_{7}$-Verbindungen mit Pyrochlor- und Weberitstruktur}},
url = {https://www.degruyter.com/document/doi/10.1515/znb-1970-1121/html},
volume = {25},
year = {1970}
}

@article{Kancko2025,
author = {Kancko, Andrej and Corr{\^{e}}a, Cinthia Antunes and Colman, Ross Harvey},
doi = {10.1016/j.jallcom.2025.180142},
issn = {09258388},
journal = {Journal of Alloys and Compounds},
month = {apr},
number = {November 2024},
pages = {180142},
title = {{Glassy disordered ground states in the frustrated pyrochlore and fluorite antiferromagnets NaCd$M_2$F$_7$ ($M$ = Ni$^{2+}$, Mn$^{2+}$)}},
url = {https://linkinghub.elsevier.com/retrieve/pii/S0925838825017001},
volume = {1024},
year = {2025}
}

@article{Shannon1976,
author = {Shannon, R. D.},
doi = {10.1107/S0567739476001551},
issn = {0567-7394},
journal = {Acta Crystallographica Section A},
month = {sep},
number = {5},
pages = {751--767},
pmid = {8378247},
title = {{Revised effective ionic radii and systematic studies of interatomic distances in halides and chalcogenides}},
url = {https://scripts.iucr.org/cgi-bin/paper?S0567739476001551},
volume = {32},
year = {1976}
}

@article{Yoon2021,
author = {Yoon, Sungwon and Lee, Wonjun and Lee, S. and Park, J. and Lee, C. H. and Choi, Y. S. and Do, S.-H. and Choi, Woo-Jae and Chen, Wei-Tin and Chou, Fangcheng and Gorbunov, D. I. and Oshima, Yugo and Ali, Anzar and Singh, Yogesh and Berlie, Adam and Watanabe, I. and Choi, Kwang-Yong},
doi = {10.1103/PhysRevMaterials.5.014411},
issn = {2475-9953},
journal = {Physical Review Materials},
month = {jan},
number = {1},
pages = {014411},
title = {{Quantum disordered state in the ${J}_{1}\text{\ensuremath{-}}{J}_{2}$ square-lattice antiferromagnet ${\mathrm{Sr}}_{2}\mathrm{Cu}({\mathrm{Te}}_{0.95}{\mathrm{W}}_{0.05}){\mathrm{O}}_{6}$}},
url = {https://link.aps.org/doi/10.1103/PhysRevMaterials.5.014411},
volume = {5},
year = {2021}
}

@article{Lee2024,
author = {Lee, Wonjun and Yoon, Sungwon and Jeon, Sungmin and Cai, Yipeng and Kojima, Kenji and Nguyen, Loi T. and Cava, R. J. and Choi, Kwang-Yong and Lee, Suheon},
doi = {10.1103/PhysRevResearch.6.023225},
issn = {2643-1564},
journal = {Physical Review Research},
month = {jun},
number = {2},
pages = {023225},
publisher = {American Physical Society},
title = {{Random singlet-like state in the dimer-based triangular antiferromagnet ${\mathrm{Ba}}_{6}{\mathrm{Y}}_{2}{\mathrm{Rh}}_{2}{\mathrm{Ti}}_{2}{\mathrm{O}}_{17\ensuremath{-}\ensuremath{\delta}}$}},
url = {https://link.aps.org/doi/10.1103/PhysRevResearch.6.023225},
volume = {6},
year = {2024}
}

@misc{Supplementary,
note = {{See Supplementary Material for details on sample synthesis, X-ray diffraction, magnetization, AC susceptibility, heat capacity, NMR and $\mu$SR results.}}
}

@article{Hossain2024,
author = {Hossain, S. M. and Rahaman, S. S. and Gujrati, H. and Bhoi, Dilip and Matsuo, A. and Kindo, K. and Kumar, M. and Majumder, M.},
doi = {10.1103/PhysRevB.110.L020406},
issn = {2469-9950},
journal = {Physical Review B},
month = {jul},
number = {2},
pages = {L020406},
publisher = {American Physical Society},
title = {{Evidence of random spin-singlet state in the three-dimensional quantum spin liquid candidate ${\mathrm{Sr}}_{3}{\mathrm{CuNb}}_{2}{\mathrm{O}}_{9}$}},
url = {https://link.aps.org/doi/10.1103/PhysRevB.110.L020406},
volume = {110},
year = {2024}
}

@article{Song2021,
author = {Song, Pengbo and Zhu, Kejia and Yang, Fan and Wei, Yuan and Zhang, Lu and Yang, Huaixin and Sheng, Xian-Lei and Qi, Yang and Ni, Jiamin and Li, Shiyan and Li, Yanchun and Cao, Guanghan and Meng, Zi Yang and Li, Wei and Shi, Youguo and Li, Shiliang},
doi = {10.1103/PhysRevB.103.L241114},
issn = {2469-9950},
journal = {Physical Review B},
month = {jun},
number = {24},
pages = {L241114},
publisher = {American Physical Society},
title = {{Evidence for the random singlet phase in the honeycomb iridate ${\mathrm{SrIr}}_{2}{\mathrm{O}}_{6}$}},
url = {https://link.aps.org/doi/10.1103/PhysRevB.103.L241114},
volume = {103},
year = {2021}
}

@article{Kimchi2018b,
archivePrefix = {arXiv},
arxivId = {1710.06860},
author = {Kimchi, Itamar and Nahum, Adam and Senthil, T.},
doi = {10.1103/PhysRevX.8.031028},
eprint = {1710.06860},
issn = {2160-3308},
journal = {Physical Review X},
month = {jul},
number = {3},
pages = {031028},
publisher = {American Physical Society},
title = {{Valence Bonds in Random Quantum Magnets: Theory and Application to ${\mathrm{YbMgGaO}}_{4}$}},
url = {https://doi.org/10.1103/PhysRevX.8.031028 https://link.aps.org/doi/10.1103/PhysRevX.8.031028},
volume = {8},
year = {2018}
}

@article{Shiroka2019,
archivePrefix = {arXiv},
arxivId = {1812.09045},
author = {Shiroka, T. and Eggenschwiler, F. and Ott, H. R. and Mesot, J.},
doi = {10.1103/PhysRevB.99.035116},
eprint = {1812.09045},
issn = {24699969},
journal = {Physical Review B},
number = {3},
pages = {1--8},
publisher = {American Physical Society},
title = {{From order to randomness: Onset and evolution of the random-singlet state in bond-disordered ${\mathrm{BaCu}}_{2}{({\mathrm{Si}}_{1\ensuremath{-}x}{\mathrm{Ge}}_{x})}_{2}{\mathrm{O}}_{7}$ spin-chain compounds}},
volume = {99},
year = {2019}
}

@article{Khatua2022,
archivePrefix = {arXiv},
arxivId = {2107.08668},
author = {Khatua, J. and Gomil{\v{s}}ek, M. and Orain, J. C. and Strydom, A. M. and Jagli{\v{c}}i{\'{c}}, Z. and Colin, C. V. and Petit, S. and Ozarowski, A. and Mangin-Thro, L. and Sethupathi, K. and Rao, M. S. Ramachandra and Zorko, A. and Khuntia, P.},
doi = {10.1038/s42005-022-00879-2},
eprint = {2107.08668},
issn = {2399-3650},
journal = {Communications Physics},
month = {apr},
number = {1},
pages = {99},
title = {{Signature of a randomness-driven spin-liquid state in a frustrated magnet}},
url = {https://www.nature.com/articles/s42005-022-00879-2},
volume = {5},
year = {2022}
}

@article{Mustonen2018,
author = {Mustonen, O. and Vasala, S. and Sadrollahi, E. and Schmidt, K. P. and Baines, C. and Walker, H. C. and Terasaki, I. and Litterst, F. J. and Baggio-Saitovitch, E. and Karppinen, M.},
doi = {10.1038/s41467-018-03435-1},
issn = {20411723},
journal = {Nature Communications},
number = {1},
pages = {7},
pmid = {29540711},
title = {{Spin-liquid-like state in a spin-$\frac{1}{2}$ square-lattice antiferromagnet perovskite induced by $d^{10}-d^{0}$ cation mixing}},
volume = {9},
year = {2018}
}

@article{Mustonen2024,
author = {Mustonen, Otto H.J. and Fogh, Ellen and Paddison, Joseph A.M. and Mangin-Thro, Lucile and Hansen, Thomas and Playford, Helen Y. and Diaz-Lopez, Maria and Babkevich, Peter and Vasala, Sami and Karppinen, Maarit and Cussen, Edmund J. and Ro̷nnow, Henrik M. and Walker, Helen C.},
doi = {10.1021/acs.chemmater.3c02535},
issn = {15205002},
journal = {Chemistry of Materials},
number = {1},
pages = {501--513},
title = {{Structure, Spin Correlations, and Magnetism of the $S = \frac{1}{2}$ Square-Lattice Antiferromagnet Sr$_2$CuTe$_{1-x}$W$_{x}$O$_{6}$ $(0\leq x \leq 1)$}},
volume = {36},
year = {2024}
}

@article{Brewer1986,
author = {Brewer, J. H. and Kreitzman, S. R. and Noakes, D. R. and Ansaldo, E. J. and Harshman, D. R. and Keitel, R.},
doi = {10.1103/PhysRevB.33.7813},
issn = {0163-1829},
journal = {Physical Review B},
month = {jun},
number = {11},
pages = {7813--7816},
title = {{Observation of muon-fluorine "hydrogen bonding" in ionic crystals}},
url = {https://link.aps.org/doi/10.1103/PhysRevB.33.7813},
volume = {33},
year = {1986}
}

@article{Cai2018,
archivePrefix = {arXiv},
arxivId = {1806.10970},
author = {Cai, Y. and Wilson, M. N. and Hallas, A. M. and Liu, L. and Frandsen, B. A. and Dunsiger, S. R. and Krizan, J. W. and Cava, R. J. and Rubel, O. and Uemura, Y. J. and Luke, G. M.},
doi = {10.1088/1361-648X/aad91c},
eprint = {1806.10970},
issn = {0953-8984},
journal = {Journal of Physics: Condensed Matter},
month = {sep},
number = {38},
pages = {385802},
pmid = {30089706},
publisher = {IOP Publishing},
title = {{$\mu$SR study of spin freezing and persistent spin dynamics in ${\text{NaCaNi}}_{2}{\mathrm{F}}_{7}$}},
url = {https://iopscience.iop.org/article/10.1088/1361-648X/aad91c},
volume = {30},
year = {2018}
}

@misc{Crysalis,
title = {{CrysAlisPro, Oxford Diffraction, Agilent Technologies UK Ltd, England}},
volume = {Version 1.171.42.51},
year = {2022},
author = {Oxford UK, Rigaku Corporation},
url = {https://rigaku.com/products/crystallography/x-ray-diffraction/crysalispro},
}

@article{Palatinus2007,
author = {Palatinus, Luk{\'{a}}{\v{s}} and Chapuis, Gervais},
doi = {10.1107/S0021889807029238},
issn = {0021-8898},
journal = {Journal of Applied Crystallography},
month = {aug},
number = {4},
pages = {786--790},
publisher = {International Union of Crystallography},
title = {{SUPERFLIP - a computer program for the solution of crystal structures by charge flipping in arbitrary dimensions}},
url = {https://journals.iucr.org/paper?S0021889807029238},
volume = {40},
year = {2007}
}

@article{Petricek2023,
author = {Pet\v{r}{\'{i}}{\v{c}}ek, V{\'{a}}clav and Palatinus, Luk{\'{a}}{\v{s}} and Pl{\'{a}}{\v{s}}il, Jakub and Du{\v{s}}ek, Michal},
doi = {10.1515/zkri-2023-0005},
issn = {2194-4946},
journal = {Zeitschrift f{\"{u}}r Kristallographie - Crystalline Materials},
month = {jul},
number = {7-8},
pages = {271--282},
title = {{Jana2020 - a new version of the crystallographic computing system Jana}},
url = {https://www.degruyter.com/document/doi/10.1515/zkri-2023-0005/html},
volume = {238},
year = {2023}
}

@article{Momma2008,
author = {Momma, Koichi and Izumi, Fujio},
doi = {10.1107/S0021889808012016},
issn = {0021-8898},
journal = {Journal of Applied Crystallography},
month = {jun},
number = {3},
pages = {653--658},
title = {{VESTA: a three-dimensional visualization system for electronic and structural analysis}},
url = {https://scripts.iucr.org/cgi-bin/paper?S0021889808012016 https://journals.iucr.org/paper?S0021889808012016},
volume = {41},
year = {2008}
}

@article{Harris1995,
author = {Harris, M. J. and Zinkin, M. P. and Zeiske, T.},
doi = {10.1103/PhysRevB.52.R707},
issn = {01631829},
journal = {Physical Review B},
number = {2},
pages = {707--711},
title = {{Magnetic excitations in a highly frustrated pyrochlore antiferromagnet}},
volume = {52},
year = {1995}
}

@article{Harris1994,
author = {Harris, M. J. and Zinkin, M. P. and Tun, Z. and Wanklyn, B. M. and Swainson, I. P.},
doi = {10.1103/PhysRevLett.73.189},
issn = {00319007},
journal = {Physical Review Letters},
number = {1},
pages = {189--192},
title = {{Magnetic structure of the spin-liquid state in a frustrated pyrochlore}},
volume = {73},
year = {1994}
}

@article{Tokiwa2021,
address = {Univ Augsburg, Ctr Elect Correlat & Magnetism, Expt Phys 6, Augsburg, Germany},
author = {Tokiwa, Yoshifumi and Bachus, Sebastian and Kavita, Kavita and Jesche, Anton and Tsirlin, Alexander A and Gegenwart, Philipp},
doi = {10.1038/s43246-021-00142-1},
issn = {2662-4443},
journal = {Communications Materials},
language = {English},
month = {apr},
number = {1},
pages = {42},
title = {{Frustrated magnet for adiabatic demagnetization cooling to milli-Kelvin temperatures}},
url = {https://www.nature.com/articles/s43246-021-00142-1},
volume = {2},
year = {2021}
}

@article{Anderson1973,
author = {Anderson, P.W.},
doi = {10.1016/0025-5408(73)90167-0},
issn = {00255408},
journal = {Materials Research Bulletin},
month = {feb},
number = {2},
pages = {153--160},
title = {{Resonating valence bonds: A new kind of insulator?}},
url = {https://linkinghub.elsevier.com/retrieve/pii/0025540873901670},
volume = {8},
year = {1973}
}

@article{Pratt2000,
author = {Pratt, F. L.},
doi = {10.1016/S0921-4526(00)00328-8},
issn = {09214526},
journal = {Physica B: Condensed Matter},
pages = {710--714},
title = {{WIMDA: A muon data analysis program for the Windows PC}},
volume = {289-290},
year = {2000}
}

@article{Arnold2014,
archivePrefix = {arXiv},
arxivId = {1407.5860},
author = {Arnold, O. and Bilheux, J. C. and Borreguero, J. M. and Buts, A. and Campbell, S. I. and Chapon, L. and Doucet, M. and Draper, N. and {Ferraz Leal}, R. and Gigg, M. A. and Lynch, V. E. and Markvardsen, A. and Mikkelson, D. J. and Mikkelson, R. L. and Miller, R. and Palmen, K. and Parker, P. and Passos, G. and Perring, T. G. and Peterson, P. F. and Ren, S. and Reuter, M. A. and Savici, A. T. and Taylor, J. W. and Taylor, R. J. and Tolchenov, R. and Zhou, W. and Zikovsky, J.},
doi = {10.1016/j.nima.2014.07.029},
eprint = {1407.5860},
issn = {01689002},
journal = {Nuclear Instruments and Methods in Physics Research, Section A: Accelerators, Spectrometers, Detectors and Associated Equipment},
pages = {156--166},
publisher = {Elsevier},
title = {{Mantid - Data analysis and visualization package for neutron scattering and $\mu$ SR experiments}},
url = {http://dx.doi.org/10.1016/j.nima.2014.07.029},
volume = {764},
year = {2014}
}

@misc{Colman2024,
author = {Colman, Ross Harvey and Kancko, Andrej and Berlie, Adam},
doi = {10.5286/ISIS.E.RB2310683},
title = {{Probing the spin dynamics of a 3$d$-pyrochlore spin-glass, NaCdCo$_2$F$_7$, and quantum spin liquid candidate, NaCdCu$_2$F$_7$, STFC ISIS Neutron and Muon Source}},
url = {https://doi.org/10.5286/ISIS.E.RB2310683},
year = {2024}
}

@article{Mugiraneza2022,
archivePrefix = {arXiv},
arxivId = {2205.07107},
author = {Mugiraneza, Sam and Hallas, Alannah M.},
doi = {10.1038/s42005-022-00853-y},
eprint = {2205.07107},
issn = {2399-3650},
journal = {Communications Physics},
month = {apr},
number = {1},
pages = {95},
publisher = {Springer US},
title = {{Tutorial: a beginner's guide to interpreting magnetic susceptibility data with the Curie-Weiss law}},
url = {https://www.nature.com/articles/s42005-022-00853-y},
volume = {5},
year = {2022}
}

@article{Wilkinson2020,
archivePrefix = {arXiv},
arxivId = {2003.02762},
author = {Wilkinson, J. M. and Blundell, S. J.},
doi = {10.1103/PhysRevLett.125.087201},
eprint = {2003.02762},
issn = {0031-9007},
journal = {Physical Review Letters},
month = {aug},
number = {8},
pages = {087201},
pmid = {32909793},
publisher = {American Physical Society},
title = {{Information and Decoherence in a Muon-Fluorine Coupled System}},
url = {https://doi.org/10.1103/PhysRevLett.125.087201 https://link.aps.org/doi/10.1103/PhysRevLett.125.087201},
volume = {125},
year = {2020}
}

@article{Pratt2006,
author = {Pratt, F. L. and Blundell, S. J. and Lancaster, T. and Baines, C. and Takagi, S.},
doi = {10.1103/PhysRevLett.96.247203},
issn = {00319007},
journal = {Physical Review Letters},
number = {24},
pages = {4--7},
title = {{Low-Temperature Spin Diffusion in a Highly Ideal $S=\frac{1}{2}$ Heisenberg Antiferromagnetic Chain Studied by Muon Spin Relaxation}},
volume = {96},
year = {2006}
}

@article{Keren2004,
author = {Keren, A. and Gardner, J. S. and Ehlers, G. and Fukaya, A. and Segal, E. and Uemura, Y. J.},
doi = {10.1103/PhysRevLett.92.107204},
issn = {00319007},
journal = {Physical Review Letters},
number = {10},
pages = {10--13},
title = {{Dynamic Properties of a Diluted Pyrochlore Cooperative Paramagnet (Tb$_p$Y$_{1-p}$)$_2$Ti$_2$O$_7$}},
volume = {92},
year = {2004}
}

@article{Kermarrec2014,
author = {Kermarrec, E. and Zorko, A. and Bert, F. and Colman, R. H. and Koteswararao, B. and Bouquet, F. and Bonville, P. and Hillier, A. and Amato, A. and {Van Tol}, J. and Ozarowski, A. and Wills, A. S. and Mendels, P.},
doi = {10.1103/PhysRevB.90.205103},
issn = {1550235X},
journal = {Physical Review B - Condensed Matter and Materials Physics},
number = {20},
pages = {1--13},
title = {{Spin dynamics and disorder effects in the $S=\frac{1}{2}$ kagome Heisenberg spin-liquid phase of kapellasite}},
volume = {90},
year = {2014}
}

@article{Lee2020,
author = {Lee, S. and Klauer, R. and Menten, J. and Lee, W. and Yoon, S. and Luetkens, H. and Lemmens, P. and M{\"{o}}ller, A. and Choi, K. Y.},
doi = {10.1103/PhysRevB.101.224420},
issn = {24699969},
journal = {Physical Review B},
number = {22},
pages = {1--7},
publisher = {American Physical Society},
title = {{Unconventional spin excitations in the $S = \frac{3}{2}$ triangular antiferromagnet RbAg$_2$Cr[VO$_4$]$_2$}},
volume = {101},
year = {2020}
}

@article{Li2016,
author = {Li, Yuesheng and Adroja, Devashibhai and Biswas, Pabitra K. and Baker, Peter J. and Zhang, Qian and Liu, Juanjuan and Tsirlin, Alexander A. and Gegenwart, Philipp and Zhang, Qingming},
doi = {10.1103/PhysRevLett.117.097201},
issn = {10797114},
journal = {Physical Review Letters},
number = {9},
pages = {1--6},
title = {{Muon Spin Relaxation Evidence for the U(1) Quantum Spin-Liquid Ground State in the Triangular Antiferromagnet YbMgGaO$_4$}},
volume = {117},
year = {2016}
}

@article{Yang2024,
author = {Yang, Yan-Xing and Jiang, Cheng-Yu and Huang, Liang-Long and Zhu, Zi-Hao and Chen, Chang-Sheng and Wu, Qiong and Ding, Zhao-Feng and Tan, Cheng and Chen, Kai-Wen and Biswas, Pabi K. and Hillier, Adrian D. and Shi, You-Guo and Liu, Cai and Wang, Le and Ye, Fei and Mei, Jia-Wei and Shu, Lei},
doi = {10.1038/s41535-024-00691-x},
issn = {2397-4648},
journal = {npj Quantum Materials},
month = {oct},
number = {1},
pages = {77},
publisher = {Springer US},
title = {{Muon spin relaxation study of spin dynamics on a Kitaev honeycomb material H$_3$LiIr$_2$O$_6$}},
url = {http://dx.doi.org/10.1038/s41535-024-00691-x https://www.nature.com/articles/s41535-024-00691-x},
volume = {9},
year = {2024}
}

@article{Kawamura2019,
archivePrefix = {arXiv},
arxivId = {1907.06176},
author = {Kawamura, Hikaru and Uematsu, Kazuki},
doi = {10.1088/1361-648X/ab400c},
eprint = {1907.06176},
issn = {1361648X},
journal = {Journal of Physics Condensed Matter},
number = {50},
pmid = {31470422},
publisher = {IOP Publishing},
title = {{Nature of the randomness-induced quantum spin liquids in two dimensions}},
volume = {31},
year = {2019}
}

@article{Sanyal2021,
archivePrefix = {arXiv},
arxivId = {2006.16987},
author = {Sanyal, Sambuddha and Damle, Kedar and Chalker, J. T. and Moessner, R.},
doi = {10.1103/PhysRevLett.127.127201},
eprint = {2006.16987},
issn = {10797114},
journal = {Physical Review Letters},
number = {12},
pages = {127201},
pmid = {34597102},
publisher = {American Physical Society},
title = {{Emergent Moments and Random Singlet Physics in a Majorana Spin Liquid}},
url = {https://doi.org/10.1103/PhysRevLett.127.127201},
volume = {127},
year = {2021}
}

@article{Liu2018,
archivePrefix = {arXiv},
arxivId = {1804.06108},
author = {Liu, Lu and Shao, Hui and Lin, Yu Cheng and Guo, Wenan and Sandvik, Anders W.},
doi = {10.1103/PhysRevX.8.041040},
eprint = {1804.06108},
issn = {21603308},
journal = {Physical Review X},
number = {4},
pages = {41040},
publisher = {American Physical Society},
title = {{Random-Singlet Phase in Disordered Two-Dimensional Quantum Magnets}},
url = {https://doi.org/10.1103/PhysRevX.8.041040},
volume = {8},
year = {2018}
}

@article{Savary2017a,
author = {Savary, Lucile and Balents, Leon},
doi = {10.1103/PhysRevLett.118.087203},
issn = {10797114},
journal = {Physical Review Letters},
number = {8},
pages = {1--5},
pmid = {28282206},
title = {{Disorder-Induced Quantum Spin Liquid in Spin Ice Pyrochlores}},
volume = {118},
year = {2017}
}

@article{Watanabe2014,
archivePrefix = {arXiv},
arxivId = {1309.6309},
author = {Watanabe, Ken and Kawamura, Hikaru and Nakano, Hiroki and Sakai, T{\^{o}}ru},
doi = {10.7566/JPSJ.83.034714},
eprint = {1309.6309},
issn = {13474073},
journal = {Journal of the Physical Society of Japan},
number = {3},
pages = {1--6},
title = {{Quantum Spin-Liquid Behavior in the Spin-$\frac{1}{2}$ Random Heisenberg Antiferromagnet on the Triangular Lattice}},
volume = {83},
year = {2014}
}

@article{Ramirez2025,
author = {Ramirez, A. P. and Syzranov, S. V.},
doi = {10.1039/D4MA00914B},
issn = {2633-5409},
journal = {Materials Advances},
number = {4},
pages = {1213--1229},
publisher = {Royal Society of Chemistry},
title = {{Short-range order and hidden energy scale in geometrically frustrated magnets}},
url = {https://xlink.rsc.org/?DOI=D4MA00914B},
volume = {6},
year = {2025}
}

@article{Sana2024,
archivePrefix = {arXiv},
arxivId = {2304.13116},
author = {Sana, B. and Barik, M. and Lee, S. and Jena, U. and Baenitz, M. and Sichelschmidt, J. and Luther, S. and K{\"{u}}hne, H. and Sethupathi, K. and Rao, M. S.Ramachandra and Choi, K. Y. and Khuntia, P.},
doi = {10.1103/PhysRevB.110.134412},
eprint = {2304.13116},
issn = {24699969},
journal = {Physical Review B},
number = {13},
pages = {134412},
publisher = {American Physical Society},
title = {{Possible realization of a randomness-driven quantum disordered state in the $S = \frac{1}{2}$ antiferromagnet Sr$_3$CuTa$_2$O$_9$}},
url = {https://doi.org/10.1103/PhysRevB.110.134412},
volume = {110},
year = {2024}
}

@article{Do2014,
author = {Do, S. H. and {Van Tol}, J. and Zhou, H. D. and Choi, K. Y.},
doi = {10.1103/PhysRevB.90.104426},
issn = {1550235X},
journal = {Physical Review B - Condensed Matter and Materials Physics},
number = {10},
pages = {1--6},
title = {{Dynamical spin-orbital correlations versus random singlets in Ba$_3$CuSb$_2$O$_9$ investigated by magnetization and electron spin resonance}},
volume = {90},
year = {2014}
}

@misc{Shimokawa2025,
  archivePrefix = {arXiv},
  arxivId = {2505.11874},
  author = {Shimokawa, Tokuro and Sabharwal, Snigdh and Shannon, Nic},
  eprint = {2505.11874},
  month = {may},
  pages = {1--13},
  title = {Can experimentally-accessible measures of entanglement distinguish quantum spin liquids from disorder-driven "random singlet" phases ?},
  url = {http://arxiv.org/abs/2505.11874},
  year = {2025}
}

@article{Pal2019,
archivePrefix = {arXiv},
arxivId = {1902.06475},
author = {Pal, Santanu and Lal, Siddhartha},
doi = {10.1103/PhysRevB.100.104421},
eprint = {1902.06475},
issn = {24699969},
journal = {Physical Review B},
number = {10},
pages = {104421},
publisher = {American Physical Society},
title = {{Magnetization plateaus of the quantum pyrochlore Heisenberg antiferromagnet}},
url = {https://doi.org/10.1103/PhysRevB.100.104421},
volume = {100},
year = {2019}
}

@article{Hagymasi2022,
author = {Hagym{\'{a}}si, Imre and Sch{\"{a}}fer, Robin and Moessner, Roderich and Luitz, David J.},
doi = {10.1103/PhysRevB.106.L060411},
issn = {2469-9950},
journal = {Physical Review B},
month = {aug},
number = {6},
pages = {L060411},
title = {{Magnetization process and ordering of the $S = \frac{1}{2}$ pyrochlore Heisenberg antiferromagnet in a magnetic field}},
url = {https://link.aps.org/doi/10.1103/PhysRevB.106.L060411},
volume = {106},
year = {2022}
}

@article{Shimizu2006,
author = {Shimizu, Y. and Miyagawa, K. and Kanoda, K. and Maesato, M. and Saito, G.},
doi = {10.1103/PhysRevB.73.140407},
issn = {1098-0121},
journal = {Physical Review B},
month = {apr},
number = {14},
pages = {140407},
title = {{Emergence of inhomogeneous moments from spin liquid in the triangular-lattice Mott insulator $\kappa$--(ET)$_2$Cu$_2$(CN)$_3$}},
url = {https://link.aps.org/doi/10.1103/PhysRevB.73.140407},
volume = {73},
year = {2006}
}

@article{Uematsu2017,
archivePrefix = {arXiv},
arxivId = {1701.02255},
author = {Uematsu, Kazuki and Kawamura, Hikaru},
doi = {10.7566/JPSJ.86.044704},
eprint = {1701.02255},
issn = {13474073},
journal = {Journal of the Physical Society of Japan},
number = {4},
pages = {1--12},
title = {{Randomness-Induced Quantum Spin Liquid Behavior in the $S = \frac{1}{2}$ Random $J_1-J_2$ Heisenberg Antiferromagnet on the Honeycomb Lattice}},
volume = {86},
year = {2017}
}

@article{Uematsu2021,
archivePrefix = {arXiv},
arxivId = {2009.08630},
author = {Uematsu, Kazuki and Hikihara, Toshiya and Kawamura, Hikaru},
doi = {10.7566/JPSJ.90.124703},
eprint = {2009.08630},
issn = {13474073},
journal = {Journal of the Physical Society of Japan},
number = {12},
pages = {1--10},
title = {{Frustration-induced quantum spin liquid behavior in the $S = \frac{1}{2}$ Random-bond Heisenberg antiferromagnet on the Zigzag Chain}},
volume = {90},
year = {2021}
}

@article{Uematsu2018,
archivePrefix = {arXiv},
arxivId = {1807.08417},
author = {Uematsu, Kazuki and Kawamura, Hikaru},
doi = {10.1103/PhysRevB.98.134427},
eprint = {1807.08417},
issn = {24699969},
journal = {Physical Review B},
number = {13},
pages = {1--12},
publisher = {American Physical Society},
title = {{Randomness-induced quantum spin liquid behavior in the $S = \frac{1}{2}$ random $J_1-J_2$ Heisenberg antiferromagnet on the square lattice}},
volume = {98},
year = {2018}
}

@article{Kawamura2014,
archivePrefix = {arXiv},
arxivId = {1408.0606},
author = {Kawamura, Hikaru and Watanabe, Ken and Shimokawa, Tokuro},
doi = {10.7566/JPSJ.83.103704},
eprint = {1408.0606},
issn = {13474073},
journal = {Journal of the Physical Society of Japan},
number = {10},
pages = {1--4},
title = {{Quantum Spin-Liquid Behavior in the Spin-$\frac{1}{2}$ Random-Bond Heisenberg Antiferromagnet on the Kagome Lattice}},
volume = {83},
year = {2014}
}

@article{Uemura1994,
author = {Uemura, Y. J. and Keren, A. and Kojima, K. and Le, L. P. and Luke, G. M. and Wu, W. D. and Ajiro, Y. and Asano, T. and Kuriyama, Y. and Mekata, M. and Kikuchi, H. and Kakurai, K.},
doi = {10.1103/PhysRevLett.73.3306},
issn = {0031-9007},
journal = {Physical Review Letters},
month = {dec},
number = {24},
pages = {3306--3309},
pmid = {10057343},
title = {{Spin Fluctuations in Frustrated Kagom\'e Lattice System SrCr$_8$Ga$_4$O$_{19}$ Studied by Muon Spin Relaxation}},
url = {https://link.aps.org/doi/10.1103/PhysRevLett.73.3306},
volume = {73},
year = {1994}
}

@article{DeVries2013,
archivePrefix = {arXiv},
arxivId = {1301.4982},
author = {de Vries, M A and Piatek, J. O. and Misek, M. and Lord, J. S. and R{\o}nnow, H. M. and Bos, J-W G},
doi = {10.1088/1367-2630/15/4/043024},
eprint = {1301.4982},
issn = {1367-2630},
journal = {New Journal of Physics},
month = {apr},
number = {4},
pages = {043024},
title = {{Low-temperature spin dynamics of a valence bond glass in Ba$_2$ YMoO$_6$}},
url = {https://iopscience.iop.org/article/10.1088/1367-2630/15/4/043024},
volume = {15},
year = {2013}
}

@article{Coelho2018,
author = {Coelho, Alan A.},
doi = {10.1107/S1600576718000183},
issn = {1600-5767},
journal = {Journal of Applied Crystallography},
month = {feb},
number = {1},
pages = {210--218},
publisher = {International Union of Crystallography},
title = {{TOPAS and TOPAS-Academic : an optimization program integrating computer algebra and crystallographic objects written in C++}},
url = {https://scripts.iucr.org/cgi-bin/paper?S1600576718000183},
volume = {51},
year = {2018}
}

\end{document}


\title{Supplementary materials: Random singlet physics in the $S=\frac{1}{2}$ pyrochlore antiferromagnet NaCdCu$_2$F$_7$}

\author{A. Kancko}
\affiliation{Charles University, Faculty of Mathematics and Physics, Department of Condensed Matter Physics, Ke Karlovu 5, 121 16, Prague, Czech Republic}
\author{H. Sakai}
\affiliation{Advanced Science Research Center, Japan Atomic Energy Agency, Tokai, Ibaraki 319-1195, Japan}
\author{C. A. Corr\^ea}
\affiliation{Institute of Physics, Czech Academy of Sciences, Na Slovance 2, 182 00, Prague, Czech Republic}
\author{P. Proschek}
\affiliation{Charles University, Faculty of Mathematics and Physics, Department of Condensed Matter Physics, Ke Karlovu 5, 121 16, Prague, Czech Republic}
\author{J. Prokle\v{s}ka}
\affiliation{Charles University, Faculty of Mathematics and Physics, Department of Condensed Matter Physics, Ke Karlovu 5, 121 16, Prague, Czech Republic}
\author{T. Haidamak}
\affiliation{Charles University, Faculty of Mathematics and Physics, Department of Condensed Matter Physics, Ke Karlovu 5, 121 16, Prague, Czech Republic}
\author{M. Uhlarz}
\affiliation{Hochfeld-Magnetlabor Dresden (HLD-EMFL), Helmholtz-Zentrum Dresden-Rossendorf (HZDR), 01328
Dresden, Germany}
\author{A. Berlie}
\affiliation{ISIS Neutron and Muon Source, Rutherford Appleton Laboratory, Science and Technology Facilities Council, Chilton, Oxfordshire OX11 0QX, U.K.}
\author{Y. Tokunaga}
\affiliation{Advanced Science Research Center, Japan Atomic Energy Agency, Tokai, Ibaraki 319-1195, Japan}
\author{R. H. Colman}
\affiliation{Charles University, Faculty of Mathematics and Physics, Department of Condensed Matter Physics, Ke Karlovu 5, 121 16, Prague, Czech Republic}

\maketitle

\section{Sample preparation \& crystal growth attempt}
Polycrystalline NaCdCu$_2$F$_7$ was prepared by mixing and grinding a stoichiometric mix of dry, high-purity precursor binary fluorides (Thermofisher Scientific 99.99\% NaF metals basis, 99.5\% CuF$_2$ anhydrous and 99.99\% CdF$_2$ metals basis). This was done inside a glovebox with an argon atmosphere to prevent absorption of air moisture that would lead to possible oxidation at elevated temperatures. The mixture was filled into a hollow tubular graphite crucible and mounted inside an optical floating zone furnace (Crystal Systems Corp. FZ-T-4000-VI-VPM-PC). An initial pre-reaction was done by rapidly melting at $\sim$ 700 $\celsius$ inside a dynamic (0.25 l/min) high-pressure (8 bar) argon atmosphere, heating the graphite at 100 mm/hr in the up-down-up direction along the whole crucible length. This was repeated several times with intermittent grindings. The graphite crucible served as a vessel suitable for accommodating highly reactive melted fluorides, as well as a purifying agent to react with possible oxide impurities present in the precursor chemicals. Lastly, single-crystal growth was attempted by grinding the pre-melted polycrystalline rod and filling it into a fresh graphite crucible, then slow melting at 2 mm/hr at $\sim$ 700 $\celsius$ in the same dynamic argon atmosphere. However, no large single crystals could be obtained. Instead, a high-purity polycrystalline NaCdCu$_2$F$_7$ rod was formed and ground into a fine powder for further measurements. Upon crushing the polycrystalline rod, small crystallites were isolated and used for structure solution using single crystal X-ray diffraction (SCXRD). To confirm that the crystal structure solved by SCXRD is representative of the bulk, a section of the recrystallized rod was crushed and checked by powder X-ray diffraction (PXRD).


\section{Powder and single crystal X-ray diffraction}

A small crystallite, selected from the crushed recrystallized rod, was mounted on a glass fiber and measured at 300 K using a Rigaku Xcalibur Gemini ultra-diffractometer, with graphite-monochromated Mo K$\alpha$ ($\lambda$ = 0.71073 \AA) radiation, and an Atlas S2 CCD detector. Diffraction data were integrated using CrysAlis Pro \cite{Crysalis} with an analytical numeric absorption correction combined with a multifaceted crystal model and an empirical absorption correction using spherical harmonics.  The structure was solved by charge flipping using Superflip \cite{Palatinus2007} and refined by full-matrix least squares in Jana2020 \cite{Petricek2023}.  Structural graphics were created using Jana2020 and Vesta \cite{Momma2008}. Supplementary crystallographic data can be obtained free of charge in the CIF deposited in the Cambridge Crystallographic Data Centre (https://www.ccdc.cam.ac.uk/structures/), deposit number 2487621.

\begin{figure}[h!]
    \centering
    \includegraphics[width=0.5\linewidth]{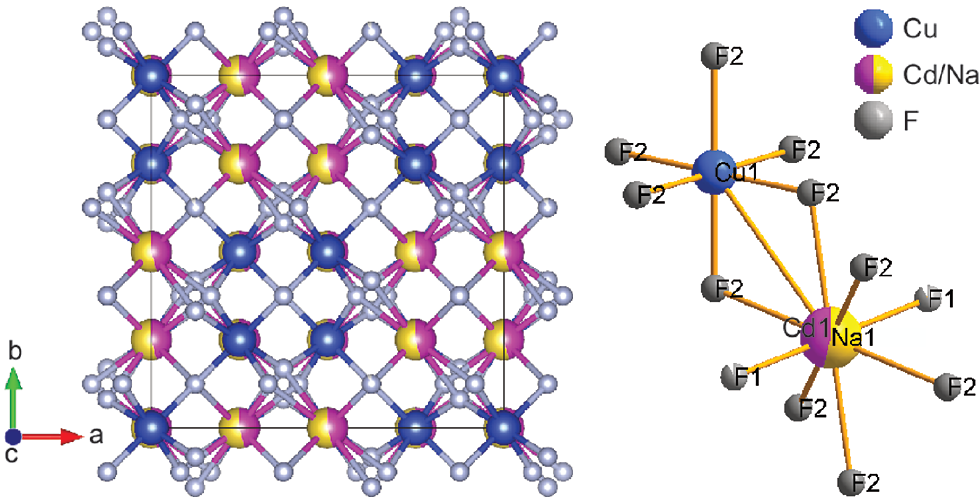}
    \caption{Unit cell of Na$_{1.076}$Cd$_{0.924}$Cu$_2$F$_{6.924}$. Na: yellow, Cd: pink, Cu: blue, F: grey.}
    \label{figS2}
\end{figure}

\begin{figure} [h!]
    \centering
    \includegraphics[width=0.65\linewidth]{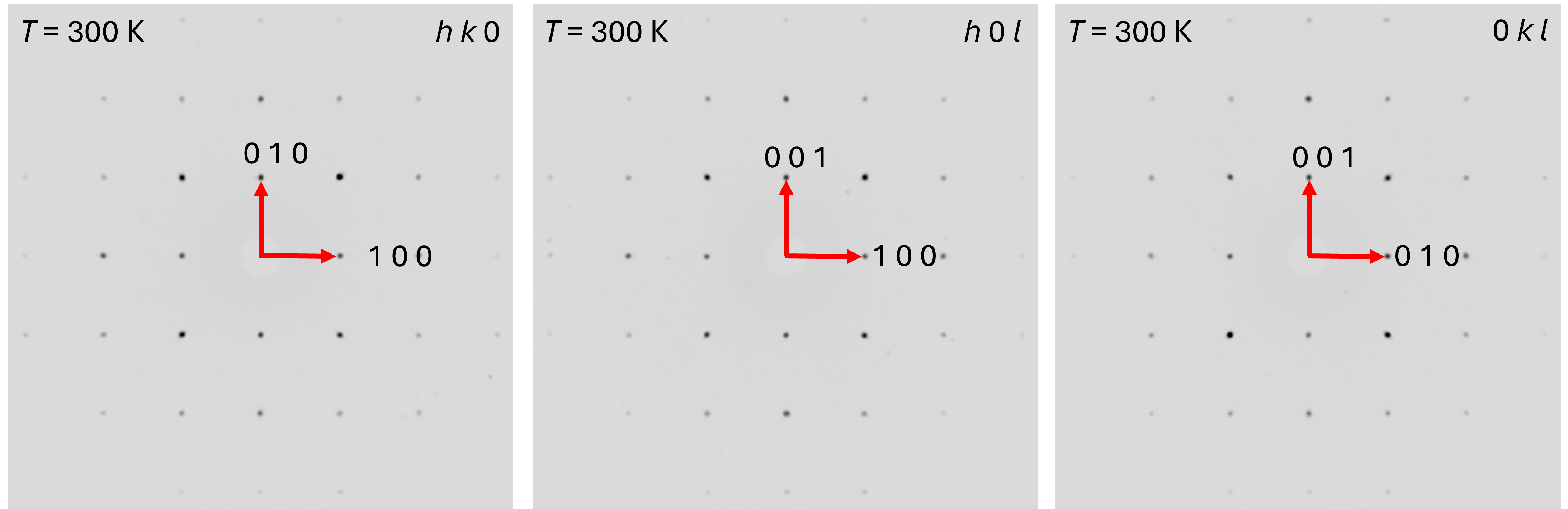}
    \caption{Reciprocal space map cuts in the $hk0$ (left), $h0l$ (middle) and $0kl$ (right) planes, obtained from SCXRD at 300 K.}
    \label{figS3}
\end{figure}

The crystal structure of Na$_{(1+x)}$Cd$_{(1-x)}$Cu$_2$F$_{(7-x)}$ is of the pyrochlore $A_2B_2X_6X'$ type and has only been reported once in a successful synthesis and powder X-ray diffraction report \cite{Hansler1970}. Similar reported structures are stoichiometric and fully fluorinated, with ions $A = A'^+_{0.5}A''^{2+}_{0.5}$ ($A'^+$ = Na, $A''^{2+}$ = Ca, Sr, Cd) sharing position with half occupancy each, and they crystallize in a cubic unit cell, space group $Fd\overline{3}m$.
The SCXRD measurement and refinement details of Na$_{1.076}$Cd$_{0.924}$Cu$_2$F$_{6.924}$ are summarized in Table \ref{tabS1}, and the structure is shown in Figure \ref{figS2}. It crystallizes in a cubic unit cell with $a$ = 10.3038(15) \AA, space group $Fd\overline{3}m$. Inspection of the reciprocal space map cuts along the $(hk0), (h0l)$ and $(0kl)$ planes reveals no superstructure reflections, which would indicate occupational ordering of Na$^+$ and Cd$^{2+}$ on the pyrochlore $A$ site (see Figure \ref{figS3}). An initial refinement of Na$_{(1+x)}$Cd$_{(1-x)}$Cu$_2$F$_{(7-x)}$ with Na and Cd occupancies fixed at 0.5 each gave residue values $R$(obs) = 2.63$\%$ and GOF(obs) = 3.12. 
Refinement of the Cd occupancy decreased the residue values to $R$(obs) = 2.19$\%$ and GOF(obs) = 1.91, which led to the conclusion that the $A$ site is Cd deficient. Chemical balance can be achieved either by having Cu$^{3+}$ sharing the $B$ site with Cu$^{2+}$ \cite{Harris1995} or F1 deficiency \cite{Harris1994}.

\begin{figure}[h!]
    \centering
    \includegraphics[width=0.4\linewidth]{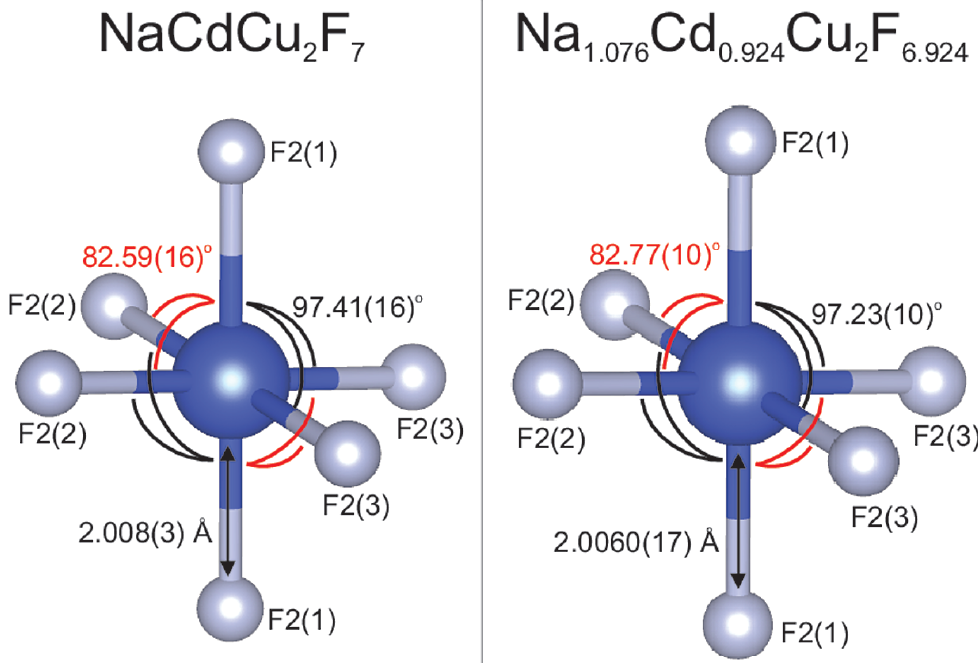}
    \caption{Cu-F2 distance and F2-Cu-F2 angles for fixed and refined occupancies of Cd and F1.}
    \label{figS4}
\end{figure}

\begin{figure}[h!]
    \centering
    \includegraphics[width=0.6\linewidth]{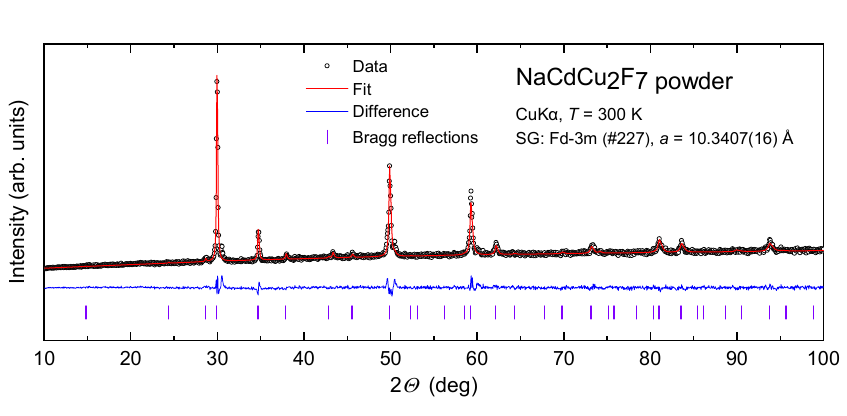}
    \caption{Powder X-ray diffraction pattern of NaCdCu$_2$F$_7$.}
    \label{figS1}
\end{figure}

According to the magnetic properties of NaCdCu$_2$F$_7$ (the experimental effective moment $\mu^{\rm exp}_{\rm eff}$ = 1.97 $\mu_B$ being close to the theoretical Cu$^{2+}$ spin-only value 1.73 $\mu_B$) and given that Cu$^{3+}$ is unlikely to be present in the sample, the occupancy of F1 was refined and resulted in $R$(obs) = 2.10$\%$ and GOF(obs) = 1.89. Na occupancy was restricted so that site $A$ has full occupancy when summing Na and Cd, while F1 was refined as a function of the Cd occupancy. Refinement of the occupancies changed neither the bond lengths, which were within 3$\sigma$ (see Figure \ref{figS4} and Table \ref{tabS2}), nor the CuF$_6$ and Na/CdF$_8$ polyhedra.
The $B$ site octahedron (CuF$_6$) is slightly deformed and has six equal bond lengths, which causes the F-Cu-F bond angle to deviate from the ideal 90$^\circ$ to 97.22(8)$^\circ$. 

A small section of the recrystallized rod obtained by the optical floating zone growth, representative of the bulk, was crushed into a fine powder and analyzed by powder X-ray diffraction (PXRD) using a Bruker D8 Advance diffractometer with CuK$\alpha_1$ and CuK$\alpha_2$ radiation ($\lambda = 1.5418$ \AA), and a structural Rietveld refinement was performed using Topas Academic V6 \cite{Coelho2018}. The PXRD pattern of NaCdCu$_2$F$_7$ is shown in Figure \ref{figS1}.

\begin{table}[h]
\centering
\caption{SCXRD structural solution and refinement data.}
\scalebox{1}{
\begin{tabular}{l|l}
\textbf{Crystal data}  \\ 
Empirical Formula & Na$_{1.076}$Cd$_{0.924}$Cu$_{2}$F$_{6.924}$ \\ 
$M_r$ & 387.2 \\ 
Crystal size (mm$^3$) & 0.137 $\times$ 0.098 $\times$ 0.033 \\ 
Crystal color and shape & Translucid irregular \\ Crystal system & Cubic \\
Cell parameter $a$ (\AA)  & 10.3038(15) \\ 
Space Group & 227 \\ 
Volume $V$ (\AA$^3$) & 1093.9(3) \\ 
Z & 8 \\ 
Density $\rho$ (calculated) (g$\cdot$cm$^{-3}$) & 4.7025 \\ 
$\mu$ (mm$^{-1}$) & 11.419 \\ 
$F$(000) & 1412 \\ \hline
\textbf{Data collection}  \\ 
Diffractometer & Xcalibur, AtlasS2, Gemini ultra \\ 
Temperature (K) & 299.82(10) \\ 
$\lambda$ (Mo K$\alpha$, average) (\AA) & 0.71073 \\ 
Data collection method & $\omega$ scans \\ 
$T$(min); $T$(max) & 0.389 ; 0.726 \\ 
$\theta$(min); $\theta$(max) ($^\circ$) & 3.42 ; 27.89 \\ 
Limiting indices & $h,k,l$ = $-13 \rightarrow 13$ \\ 
Measured / independent reflections & 3117 / 85 \\ 
$R_{\rm int}$ ($\%$) & 5.16 \\ 
Absorption Correction & Analytical \\  \hline
\textbf{Refinement}  \\
Software & Jana2020 \\ 
Solution method & Charge flipping \\ 
ADP & Anisotropic \\ 
Number of parameters refined & 12 \\ 
Restrains & 1 \\ 
R(F$^2$ $>$ 3$\sigma$(F$^2$)) / R(all) ($\%$) & 2.10 / 3.34 \\ 
wR(F$^2$ $>$ 3$\sigma$(F$^2$)) / wR(all) ($\%$) & 6.15 / 6.77 \\
Weighting scheme & $w = 1 / [ \sigma^2(F_o^2) + (0.02P)^2 + 0.0000P]$, $P = (F_o^2 + 2F_c^2) / 3$ \\ 
GOF(obs) / GOF(all) & 1.89 / 1.75 \\ 
Fourier difference residues (min ; max) (e\AA$^{-3}$) & $-0.45$ ; $0.54$ \\ 
\end{tabular}}
\label{tabS1}
\end{table}

\begin{table}[h]
\caption{F2--Cu--F2 angles and Cu--F2 distance for fixed and refined occupancies of Cd and F1.}
\renewcommand{\arraystretch}{1.5}
\setlength{\tabcolsep}{4pt}
\begin{tabular}{ccc}
                    \hline & Ideal composition & Refined composition     \\ 
                     & NaCdCu$_2$F$_7$         & Na$_{1.076}$Cd$_{0.924}$Cu$_2$F$_{6.924}$ \\ \hline
F2(1)--Cu--F2(2) angle & 82.59(16)$^{\circ}$          & 82.77(10)$^{\circ}$              \\
F2(1)--Cu--F2(3) angle & 97.41(16)$^{\circ}$         & 97.23(10)$^{\circ}$               \\
F2(1)--Cu distance    & 2.008(3) \AA        & 2.0060(17) \AA            
\end{tabular}
\label{tabS2}
\end{table}

\clearpage


\section{DC Magnetization}

DC magnetization measurements on powdered NaCdCu$_2$F$_7$ were performed in a Quantum Design Magnetic Property Measurement System (model MPMS-XL 7 T), using the reciprocating sample option (RSO). 

\begin{figure}[h]
    \centering
    \includegraphics[width=0.8\linewidth]{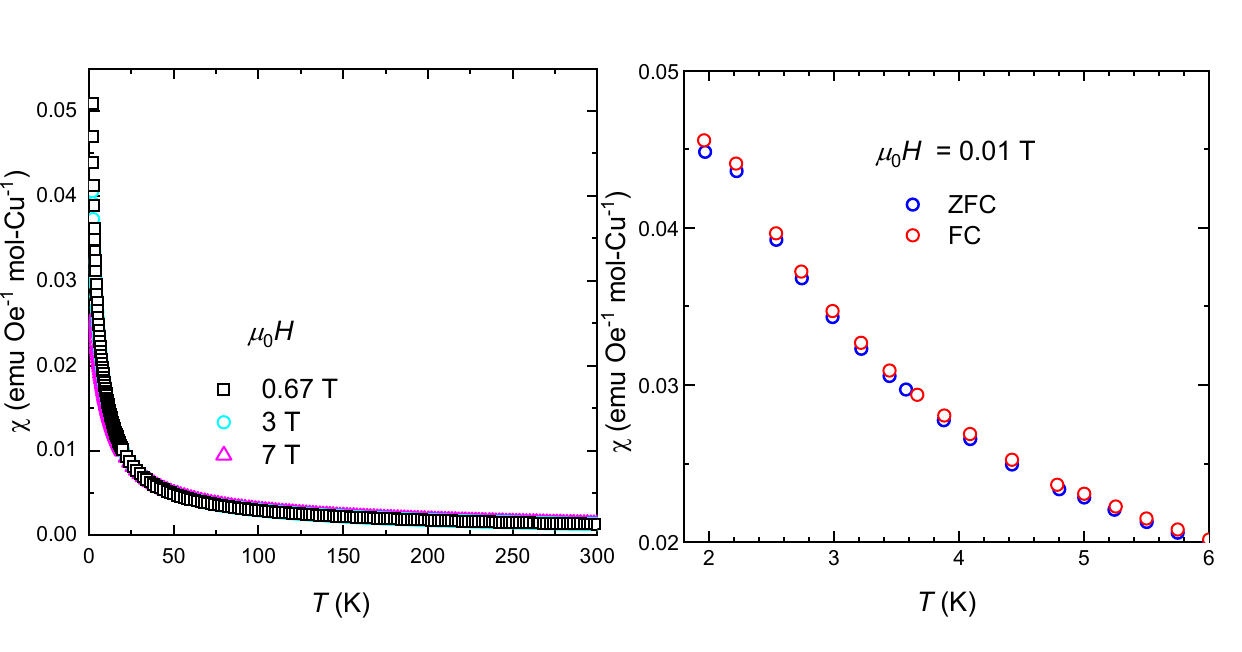}
    \caption{Magnetic susceptibility $\chi(T)=M/H$ in 0.67 T, 3 T and 7 T fields (left panel). Zero-field-cooled (ZFC) and field-cooled (FC) $\chi(T)$ in a 100 Oe (0.01 T) field  (right panel).}
    \label{figS5}
\end{figure}

In the left panel of Figure \ref{figS5}, we show the DC magnetic susceptibility $\chi(T) = M(T)/H$ measured in $\mu_0H =$ 0.67 T, 3 T and 7 T fields between 1.8 and 300 K. As discussed in the main text, we observe a diverging Curie-like tail at low temperature coming about from the paramagnetic orphan spins in the material. With applied field, the low-temperature tail gets supressed and begins to show curvature as we saturate the orphan spins contribution. No magnetic transition or spin freezing is observed in $\chi(T)$ down to 1.8 K. This is proven by measuring the zero-field-cooled (ZFC) and field-cooled (FC) susceptibility in a small applied field of 100 Oe (0.01 T) -- we see no bifurcation of $\chi_{\rm ZFC}(T)$ and $\chi_{\rm FC}(T)$, indicating a dynamic state without any history dependence (right panel of Figure \ref{figS5}).  

Isothermal magnetization $M(H)$ was measured between 1.8 and 20 K in fields up to 7 T. No hysteresis effects were observed, instead we observe a superposition of a Brillouin contribution from weakly interacting orphan spins and an intrinsic antiferromagnetic response from the random singlet phase. As discussed in the main text, the data were fitted using a modified $S=\frac{1}{2}$ Brillouin function and a power-law term:

\begin{equation}
 M(H) = \frac{n_\mathrm{orp} g \mu_B}{2} \mathrm{tanh} \left(\frac{g \mu_B \mu_0 H}{2k_\mathrm{B} (T-T^*)}\right) + A(\mu_0 H)^\alpha
 \label{eq1supp}
\end{equation}

Initially, a fit was performed on the 1.8 K data, yielding a $n_{\rm orp}$ = 15.4(2) \% fraction of weakly ferromagnetically coupled ($T^*$ = 0.30(2) K) orphan spins, and an exponent $\alpha_M$ = 0.728(15) for the intinsic power-law response. The orphan spins fraction $n_{\rm orp}$ and the interaction temperature $T^*$ were then fixed and a global fit was performed on $M(H)$ for temperatures between 2 and 20 K. The fits are shown in the main text in Figure 2c , with a decomposition of the 1.8 K fit into the Brillouin and intrinsic component in Figure 2d. The fitted parameters are summarized in Table \ref{tab3supp}. 

\begin{figure}
    \centering
    \includegraphics[width=0.85\linewidth]{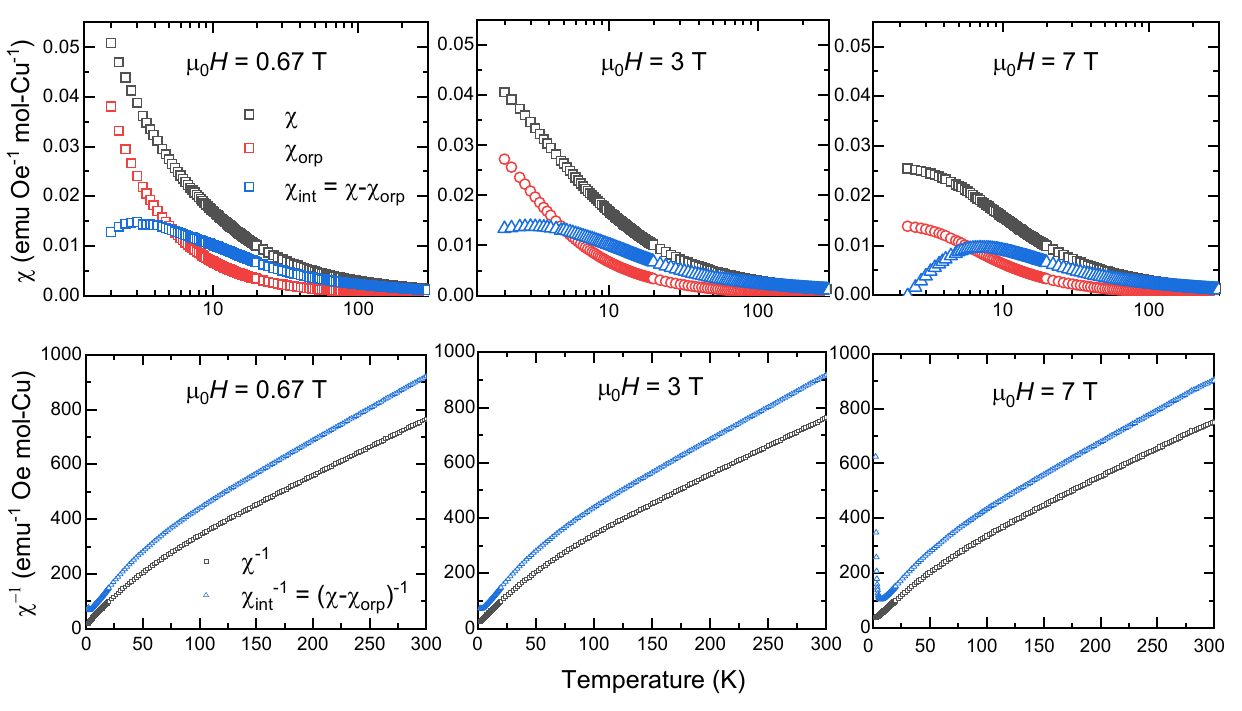}
    \caption{Decomposition of magnetic susceptibility $\chi(T)$ in different fields into the orphan spins contribution $\chi_{\rm orp}$ and the intrinsic susceptibility $\chi_{\rm int} = \chi - \chi_{\rm orp}$ (top panel). Inverse susceptibility $\chi^{-1}(T)$ and  $\chi_{\rm int}^{-1}(T)$ in different fields (bottom panel).}
    \label{figS6}
\end{figure}

The orphan spins contribution $\chi_{\rm orp}$ was then subtracted from the magnetic susceptibility $\chi(T)$ at 0.67 T, 3 T and 7 T, resulting in the intrinsic susceptibility $\chi_{\rm int} = \chi - \chi_{\rm orp}$ (see Figure \ref{figS6}). At 0.67 T we notice a broad maximum in $\chi_{\rm int}$ around 3 K, which gets suppressed and shifts to higher temperature with applied field. In the bottom panel of Figure \ref{figS6} we plot the inverse susceptibility $\chi^{-1}(T)$ and $\chi_{\rm int}^{-1}(T)$. Performing the Curie-Weiss fit on the inverse intrinsic susceptibility $\chi_{\rm int}^{-1} = (\chi - \chi_{\rm orp})^{-1}$ at 0.67 T between 200 and 300 K, we extract $C$ = 0.427(1) emu K mol$^{-1}$ Oe$^{-1}$, $\theta_{\mathrm{CW}} = -93(1)$ K and $\mu_{\mathrm{eff}}$ = 1.85(1) $\mu_B$, corresponding to $g = 2.14(1)$.

\begin{table}[htbp]
\caption{Fitted parameters from Equation \ref{eq1supp}.}
\centering
\small
\setlength{\tabcolsep}{4pt}
\renewcommand{\arraystretch}{1.3}
\begin{tabular}{c|ccccccccc}
\midrule
$T$ (K) & 1.8 & 2 & 3 & 4 & 6 & 8 & 10 & 20   \\
\midrule
$n_{\rm orp}$         & \multicolumn{9}{c}{0.154(2)} \\
$g$         & \multicolumn{9}{c}{2}  \\
$\alpha_{M}$           & 0.728(15) & 0.708(13) & 0.685(7) & 0.727(6) & 0.817(7) & 0.878(8) & 0.914(9) & 0.985(14)  \\
$T^*$ (K)    & \multicolumn{9}{c}{0.30(2)} \\
$A$ ($\mu_{\rm B}$ T$^{-\alpha_M}$) & 0.028(1) & 0.029(1) & 0.030(1) & 0.028(1) & 0.023(1) & 0.019(1) & 0.016(1) & 0.010(1) & \\
\end{tabular}
\label{tab3supp}
\end{table}

High-field isothermal magnetization was measured in the High Magnetic Field Laboratory (Hochfeld-Magnetlabor Dresden, HLD) in the Helmholtz-Zentrum Dresden-Rossendorf (HZDR) in Dresden, Germany. A pulsed 60 T magnet was utilized, with a cryostat setup allowing us to measure $M$($H$) between 1.35 and 50 K. The powdered NaCdCu$_2$F$_7$ was stuffed inside a kapton tube, which was attached to a sample holder and
inserted into the magnetometer inside the cryostat. The magnetometer consists of two pick-up coils, balanced carefully to subtract the
magnet-pulse background from the sample signal. The high-field magnetization data measured on sweeping up the field are shown in the main text in Figure 2e.

\section{AC susceptibility}

Temperature and frequency dependent AC susceptibility was measured between 0.4 and 4 K using a custom-made coil for the $^3$He option in a Quantum Design Physical Property Measurement System (PPMS). An oscillating field with a $H_{\rm AC}$ = 0.05 Oe amplitude was used, with no additional external applied field. The used frequency range was 10 -- 10 000 Oe, equidistant on a logarithmic scale. To acquire reasonable statistics, frequencies between 10 and 464.2 Hz were measured with 5000 repetitions, and frequencies between 1000 and 10 000 Hz were measured using a 5 second counting time. The data are summarized in the main text in Figure 2f. Frequencies below 100 Hz have been ommitted due to a large noise contribution.

\section{Heat capacity}
Heat capacity was measured between 1.8 K and 300 K via the heat relaxation method in the PPMS, with the addition of a $^3$He low-temperature insert for temperatures between 0.4 and 20 K. A Quantum Design heat capacity puck with a non-magnetic sapphire stage was utilized, using a small amount of Apiezon-N grease for attaching the polished piece of polycrystalline NaCdCu$_2$F$_7$ to the platform. The apiezon contribution to specific heat was subsequently removed. An Oxford Instruments Triton dilution refrigerator was employed for zero-field and in-field sub-1 K heat capacity measurements. 
For the estimation of the magnetic component of specific heat, the subtraction of the phonon contribution from the lattice was performed using a non-magnetic analogue NaCdZn$_2$F$_7$ (similarly as for NaCdCo$_2$F$_7$, NaCdNi$_2$F$_7$ and NaCdMn$_2$F$_7$ \cite{Kancko2023, Kancko2025}), with some scaling necessary to accommodate for the differing unit cell volume: 
\begin{equation}
C_{\rm mag} = C_p - C_{\rm lat} \approx C_{p \rm[NaCdCu_2F_7]} - C_{p \rm[NaCdZn_2F_7]}  
\label{eq2supp}
\end{equation}

\begin{figure}
    \centering
    \includegraphics[width=0.9\linewidth]{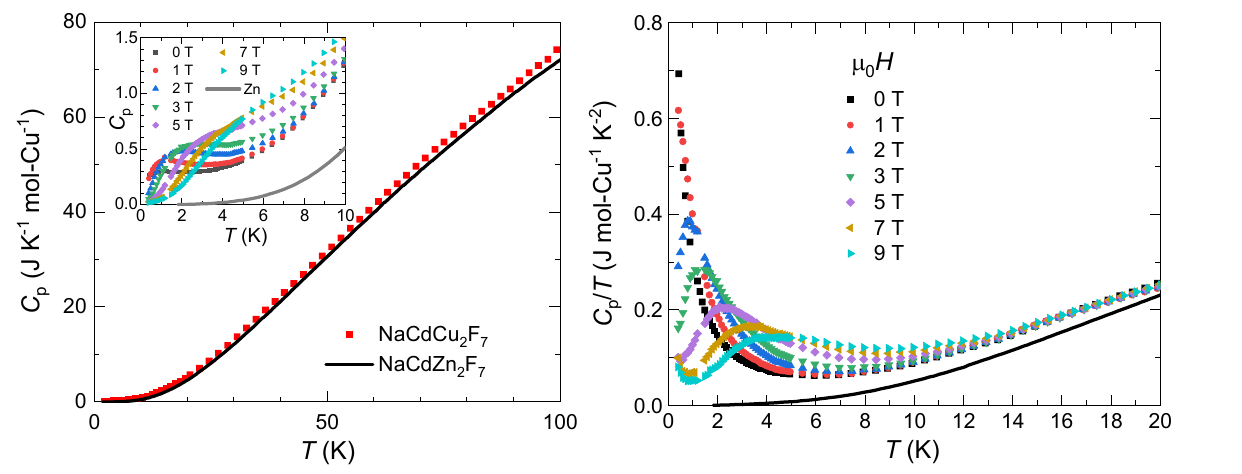}
    \caption{Specific heat $C_p$ of NaCdCu$_2$F$_7$ and the non-magnetic analogue NaCdZn$_2$F$_7$ as a solid line (left panel), with the 0.4 -- 20 K region zoomed in with applied field up to 9 T (left inset). Evolution of specific heat divided by temperature $C_p/T$ in applied field (right panel).}
    \label{fig9supp}
\end{figure}

In the left panel of Figure \ref{fig9supp}, we see the zero field specific heat $C_p$ of NaCdCu$_2$F$_7$ between 1.8 and 100 K, with the scaled $C_p$ of the non-magnetic analogue NaCdZn$_2$F$_7$ (black solid line) for eventual subtraction of phonon contribution. In the inset, we zoom in to the $^3$He specific heat measured between 0.4 and 20 K in fields up to 9 T. We see a broad anomaly centered around 1 K, which shifts to higher temperature with applied field. In the right panel of Figure \ref{fig9supp}, we show the specific heat divided by temperature $C_p/T$ in different fields. We observe an emerging peak in $C_p/T$ below 1 K, which flattens and shifts to higher temperature with applied field. Both the zero-field and in-field data overlap above 15 K, suggesting a dominant phonon contribution above this temperature. 

The extracted magnetic specific heat $C_{\rm mag}$ and $C_{\rm mag}/T$ are shown in the main text in Figure 3c and 3d. We observe a broad peak shifting to higher temperature with applied field, which comes from a combination of a Schottky anomaly of the $S=\frac{1}{2}$ orphan spins and an intrinsic random-singlet term. We can fit the data using a two-level Schottky anomaly and a power-law term corresponding to the random singlet phase \cite{Kimchi2018,Hossain2024}:

\begin{equation}
    C_{\rm mag} = C_{\rm Sch} + C_{\rm power-law} 
    \label{eq3supp}
\end{equation}

\begin{figure}
    \centering
    \includegraphics[width=0.85\linewidth]{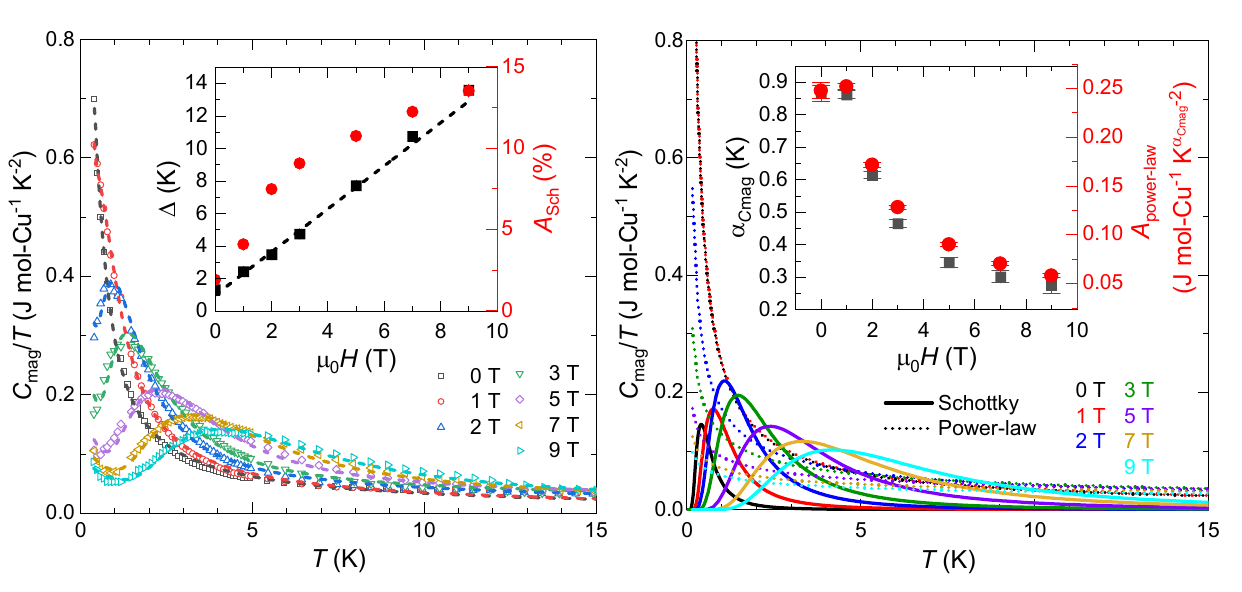}
    \caption{Magnetic specific heat $C_{\rm mag}/T$ with fits to Equation \ref{eq4supp} shown as dashed lines (left panel). The left inset shows the field dependence of the Schottky gap $\Delta(H)$ (left axis, black) with a linear fit to the Zeeman splitting $\Delta(H) = gS\mu_B\mu_0H$ (dashed line), giving $gS$ = 1.32(6) or $g$ = 2.6(1) for $S=\frac{1}{2}$. The Schottky scaling parameter $A_{\rm Sch}$ is shown on the right axis of the left inset (red). The right panel shows the individual fitting curves - power-law term as dotted lines and Schottky term as solid lines. The right inset shows the field dependence of the power-law exponent $\alpha_{C_{mag}}$ (left axis, black) and the scaling parameter $A_{\rm power-law}$ (right axis, red).}
    \label{fig11supp}
\end{figure}

Taking the standard formula for a two-level Schottky system and $C_{\rm power-law} = A_{\rm power-law}T^{1-\alpha_{C_{\rm mag}}}$, we then fit the $C_{\rm mag}/T$ data as follows:

\begin{equation}
    \frac{C_{\rm mag}}{T} = A_{\rm Sch}\mathrm{R} \frac{\Delta^2}{T^3} \frac{\mathrm{exp}\left(\frac{\Delta}{T}\right)}{\left(1+\mathrm{exp}\left(\frac{\Delta}{T}\right) \right)^2} + A_{\rm power-law}T^{-\alpha_{C_{\rm mag}}}
    \label{eq4supp}
\end{equation}

where $\Delta$ is the energy gap and $\alpha_{C_{\rm mag}}$ is the power-law exponent. The fits are shown in Figure \ref{fig11supp}, with the fitted parameters summarized in Table \ref{tab4supp}. In the left inset of Figure \ref{fig11supp}, we see the field dependence of the Schottky gap $\Delta(H)$ as well as the scaling parameter $A_{\rm Sch}$. We fit the field dependence of the Schottky gap with the Zeeman splitting formula $\Delta(H) = gS\mu_B\mu_0H$ and extract the slope $gS$ = 1.32(6), corresponding to the $g$-factor for the $S=\frac{1}{2}$ orphan spins $g$ = 2.6(1). In the right panel of Figure \ref{fig11supp}, we see the field dependence of the exponent $\alpha_{\rm power-law}$ and the scaling factor $A_{\rm power-law}$. Both parameters decrease with applied field, the power-law exponent changing from 0.87(2) in zero field to 0.28(2) at 9 T. 

By integrating the Schottky contribution $\Delta S_{\rm Sch} = \int_{0.4}^{20} \frac{C_{\rm Sch}}{T} dT$ and the magnetic specific heat $\Delta S_{\rm mag} = \int_{0.4}^{20} \frac{C_{\rm mag}}{T} dT$ between 0.4 and 20 K, we get the respective magnetic entropy changes. We can estimate the fraction of orphan spins by taking the ratio $n_{\rm orp} = \frac{\Delta S_{\rm Sch}}{\Delta S_{\rm mag}}$. In Table \ref{tab4supp}, we see that the orphan spins fraction $n_{\rm orp}$ changes from $\sim 10\%$ in zero field to $\sim 54~\%$ in 9 T field.

\begin{table}[h]
\caption{Summary of fitted parameters from Equation \ref{eq4supp}.}
\setlength{\tabcolsep}{4pt}
\renewcommand{\arraystretch}{1.3}
\begin{tabular}{c|ccccccc}
\hline
$\mu_0 H$ (T)                         & 0                 & 1                 & 2                 & 3                 & 5                 & 7                  & 9                  \\ \hline
$\Delta$ (K)                          & 1.34(5) & 2.44(5) & 3.49(3) & 4.76(4) & 7.74(8) & 10.76(12) & 13.60(17) \\
$A_{\rm Sch}$                         & 0.019(3) & 0.041(1)  & 0.075(1)  & 0.091(1)  & 0.107(2) & 0.122(3)  & 0.135(4)  \\
$\alpha_{C_{mag}}$                      & 0.87(2) & 0.86(1) & 0.61(1) & 0.47(1) & 0.35(1) & 0.30(2)   & 0.28(2) \\
$A_{\rm power-law}$ (J mol-Cu$^{-1}$ K$^{\alpha-2}$)& 0.248(8) & 0.252(3) & 0.172(2) & 0.128(2) & 0.090(2) & 0.071(2)   & 0.058(2)  \\ \hline
$\Delta S_{\rm Sch}$ (J mol-Cu$^{-1}$ K$^{-2}$) & 0.10964 &  0.23424 & 0.42688
& 0.51273 & 0.59052 & 0.64366 & 0.67455 \\ 
$\Delta S_{\rm mag}$ (J mol-Cu$^{-1}$ K$^{-2}$) & 1.1379 & 1.30967 &  1.3551  & 1.36089  & 1.34662  & 1.29536 & 1.24204 \\
$n_{\rm orp}$ = $\frac{\Delta S_{\rm Sch}}{\Delta S_{\rm mag}}$ & 0.09635 & 0.17885 & 0.31502 & 0.37676 & 0.43852 & 0.4969 & 0.5431

\end{tabular}
\label{tab4supp}
\end{table}

\section{Muon spin relaxation ($\mu$SR)}
Muon spin relaxation spectroscopy measurements were performed at the ISIS Neutron and Muon Source at the Rutherford Appleton Laboratory in Harwell Oxford, United Kingdom. The MuSR instrument was used, utilising a dilution refrigerator (50 mK - 4 K) cryostat setup. Approximately 600 mg of powdered phase-pure NaCdCu$_2$F$_7$ was spread onto a silver plate, with a small addition of a dilute GE varnish to stick the thin layer of powder to the surface, and enclosed with a second Ag plate from the top. The sample-containing Ag block was then attached to the sample stick and inserted into the dilution refrigerator cryostat in the MuSR instrument. A TF20 (20 Gauss transverse field) measurement was performed at 10 K to calibrate the asymmetry parameter between the front and back detectors.  The sample was then cooled down to the base temperature (50 mK), where the isothermal field dependence (0 - 3000 G) of the muon asymmetry spectra was taken. Afterwards, the temperature dependence (50 mK - 4 K) of both zero-field and longitudinal-field muon asymmetry spectra were collected, see Figure 4 in the main text.  The spectra were fitted using the WiMDA and Mantid software. \cite{Pratt2000,Arnold2014}

\begin{figure}
    \centering
    \includegraphics[width=0.85\linewidth]{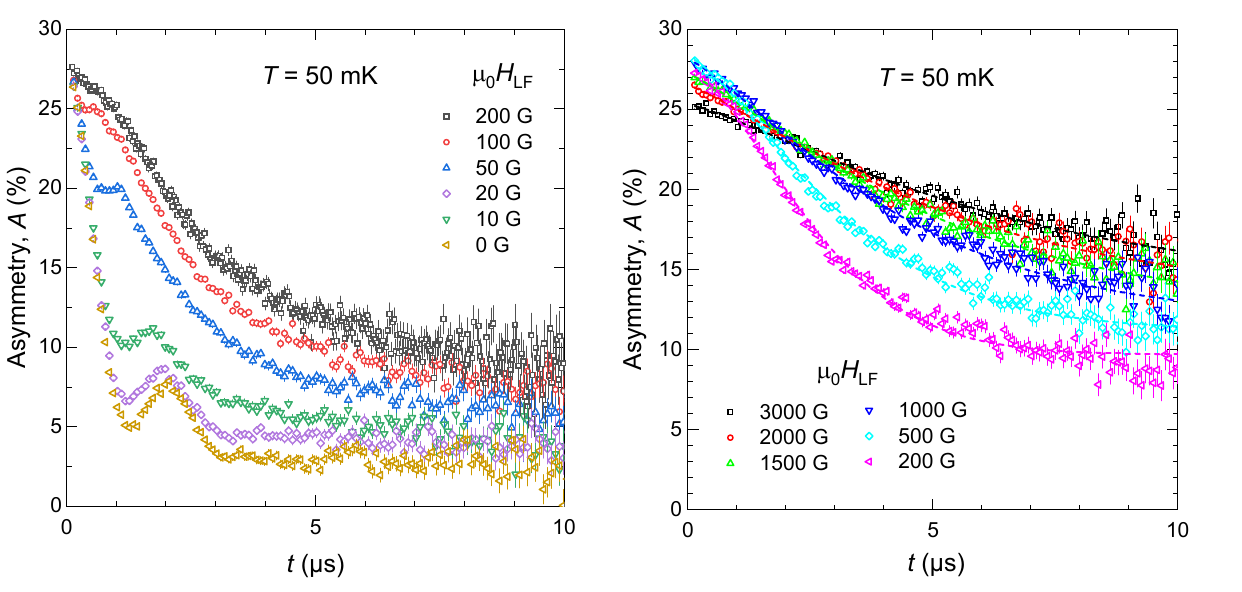}
    \caption{Field dependence of $\mu$SR spectra at the base temperature 50 mK between 0 and 200 G (left panel). Field dependence of the 50 mK spectra between 200 and 3000 G (right panel).}
    \label{fig14supp}
\end{figure}

\begin{figure}
    \centering
    \includegraphics[width=0.85\linewidth]{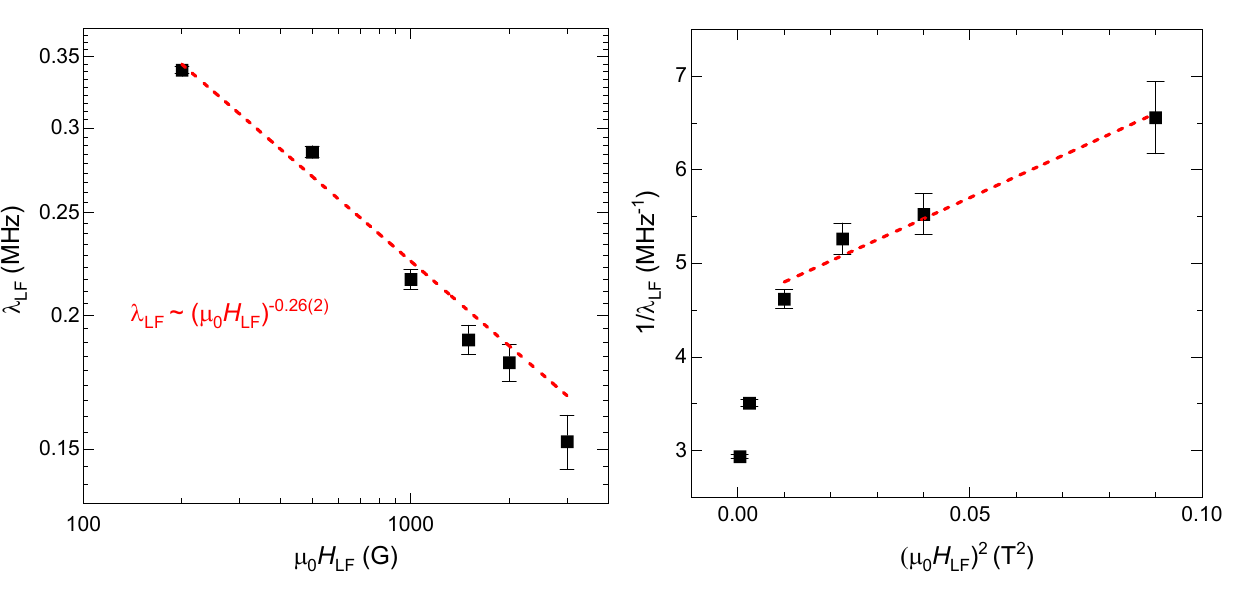}
    \caption{Field dependence of the fitted muon relaxation rate $\lambda$ vs. $H_{\rm LF}$ on a log-log scale, with a linear fit shown as the red dashed line, giving a power-law dependence $\lambda \sim (\mu_0 H_{\rm LF})^{{\alpha_{\mu \rm SR}}-1}$ with $\alpha_{\mu \rm SR}$ = 0.74(2) (left panel). The right panel shows the $1/\lambda$ vs. $(\mu_0 H_{\rm LF})^2$ dependence with the BPP fit between 1000 and 3000 G shown as the red dashed line.}
    \label{fig15supp}
\end{figure}

The base-temperature (50 mK) isothermal field dependence of $\mu$SR spectra can be seen in Figure \ref{fig14supp}. In the left panel, we see the 0--200 G field dependence. We notice oscillations coming from the F-$\mu$-F complex, as expected in all fluoride materials \cite{Brewer1986}. The static fields coming from the F-$\mu$-F dipole-dipole interaction can be decoupled by applying a small longitudinal field (LF), as we suppress the oscillations to a shorter time window with increasing field, eventually fully decoupling this effect with the application of LF 200 G. In the right panel of Figure \ref{fig14supp}, we see the isothermal field dependence at 50 mK between 200 and 3000 G after fully decoupling the F-$\mu$-F oscillations. We observe a Gaussian shoulder below $t <$ 3 $\mu$s for fields between 200 and 1000 G, suggesting the presence of some static internal fields likely from a frozen subset of orphan spins or dimer singlets. As we increase the field to 1500 G and above, we fully decouple the static internal fields and observe a fully exponential relaxation of asymmetry, suggesting a fully dynamic state. We fit the spectra using a stretched exponential function $A(t) = A_0 \mathrm{exp}(-(\lambda_{\rm LF} t)^\beta) + A_{\rm base}$, where $A_0$ is the relaxing amplitude, $\lambda_{\rm LF}$ is the muon relaxation rate, $\beta$ is the stretching exponent and $A_{\rm base}$ is the non-relaxing constant baseline. In Figure \ref{fig15supp}, we plot the muon relaxation rate $\lambda_{\rm LF}$ as a function of the applied field $\mu_0 H_{\rm LF}$ on a log-log scale (left panel), as well as the inverse $1/\lambda_{\rm LF}$ as a function of $H^2_{\rm LF}$ on a linear scale. We observe a power-law decay of muon relaxation rate $\lambda_{\rm LF} \sim (\mu_0 H_{\rm LF})^{{\alpha_{\mu \rm SR}}-1}$ with $\alpha_{\mu \rm SR}$ = 0.74(2). This leads to a data collapse of the field-dependent muon spectra when plotted as polarization $P(t) = (A(t)-A_{\rm base})/A_0$ as a function of time scaled by field, $t/(\mu_0 H)^{\eta_{\mu \rm SR}}$, for $\eta_{\mu \rm SR} = 1-\alpha_{\mu \rm SR} = 0.26(2)$, see Figure 6e in the main text. 

Furthermore, the $(\mu_0 H_{\rm LF})^2$ dependence of $1/\lambda_{\rm LF}$ (right panel of Figure \ref{fig15supp}) can be seen for fields between 1000 and 3000 G. The high-field regime can be analyzed using the Bloembergen Purcell and Pound (BPP) theory, which assumes simple spin dynamics with an exponentially decaying spin-spin autocorrelation function $\langle S(t)S(0) \rangle \sim e^{-\nu t}$, for which: 
\begin{equation}
    \lambda_{\rm LF} = \frac{2 \gamma_\mu^2 H^2_\mathrm{fluc} \nu}{\nu^2 + \gamma_\mu^2 H_{\mathrm{LF}}^2}
    \label{eq5supp}
\end{equation}
where $\gamma_\mu$ = 135.538 MHz/T is the muon gyromagnetic ratio, $H_\mathrm{fluc}$ is the internal fluctuating field and $\nu$ is the fluctuating frequency. From our BPP fit for fields between 1500 and 3000 G, we get $H_\mathrm{fluc}$ = 191(7) G and $\nu$ = 61(5) MHz. 

\section{Nuclear magnetic resonance (NMR)}
NMR measurements in the temperature range of 1.8 - 150 K were performed  using a $^4$He variable-temperature insert (VTI) installed in a superconducting magnet dedicated to NMR, together with a phase-coherent, pulsed spectrometer.
For measurements between 120 mK and 2.1 K, a $^3$He-$^4$He dilution-refrigerator probe head was operated in the VTI.
The excitation coil was custom-wound from silver wire.
Frequency-swept spectra at different constant fields were obtained with the radio frequency (r.f.) circuit tuned and matched at each point.
Nuclear spin echoes were generated using a standard 90$\degree$ - 180$\degree$ pulse sequence, with a first-pulse duration of 2-3 $\mu$s for the $^4$He cryostat range and 10-20 $\mu$s for the dilution cryostat range; the r.f. power was optimized for each spectral peak. The interpulse delay $\tau$ was typically 20-30 $\mu$s.
Each echo signal was accumulated and Fourier transformed to obtain the spectra.   

For every temperature, the nuclear spin-lattice relaxation rate $1/T_1$ was measured at the position of the resonance peak by fitting the recovery curves with a stretched exponential function:
\begin{equation}
    \frac{M(\infty)-M(t)}{M(\infty)} = 0.4 \exp \{-\left( t/T_1 \right)^{\beta_{\rm NMR}} \} + 0.6 \exp \{ -\left( 6t/T_1 \right)^{\beta_{\rm NMR}} \},
\end{equation} 

assuming that only the central transition was saturated while the populations of satellite transitions remain unaffected. The stretching exponent $\beta_{\rm NMR}$ is indicative of a spatial distribution of $T_1$ values. The temperature dependence of the extracted $1/T_1$ and $\beta$ values is summarized in Figure 5 in the main text..


\bibliography{bibliography}